\documentclass[12pt]{article}
\usepackage{amssymb, graphicx, hyperref}
\begin{document}
\title{\bf\Large Symmetries, Matrices, and de Sitter Gravity\vspace{18pt}}
\author{\normalsize 
Yi-hong Gao\vspace{12pt}\\ 
{\it\small Institute of Theoretical Physics, Beijing 100080, 
            P.~R.~China}
\\ {\small e-mail: {\tt gaoyh@itp.ac.cn}}}
\date{}
\maketitle
\baselineskip .64cm
\renewcommand{\theequation}{\thesection.\arabic{equation}}
\csname @addtoreset\endcsname{equation}{section}
\voffset -.2in
\vskip 1cm
\centerline{\bf Abstract}
\vskip .4cm
Using simple algebraic methods along with an analogy to the BFSS model, we explore the possible (target) spacetime symmetries that may appear in a matrix description of de Sitter gravity. Such symmetry groups could arise in two ways, one from an ``IMF'' like construction and the other from a ``DLCQ'' like construction. In contrast to the flat space case, we show that the two constructions will lead to different groups, {\it i.e.} the Newton-Hooke group and the inhomogeneous Euclidean group (or its algebraic cousins). It is argued that matrix quantum mechanics based on the former symmetries look more plausible. Then, after giving a detailed description of the relevant one particle dynamics, a concrete Newton-Hooke matrix model is proposed. The model naturally incorporates issues such as holography, UV-IR relations, and fuzziness, for gravity in $dS_{4}$. We also provide evidence to support a possible phase transition. The lower temperature phase, which corresponds to gravity in the perturbative regime, has a Hilbert space of infinite dimension. In the higher temperature phase where the perturbation theory breaks down, the dimension of the Hilbert space may become finite. 
\newpage
\section{Introduction}
Quantum gravity in de Sitter space has attracted some recent interest \cite{tb}--\cite{Klemm} (see also \cite{de} for some earlier discussions). At present, there are basically two kinds of proposals in search for a dual description of de Sitter gravity, --- one mimics the AdS/CFT correspondence \cite{Mal} and the other attempts to model on Matrix Quantum Mechanics of M-theory in flat space background \cite{BFSS}. A common motivation of these investigations is the holographic principle, which tries to associate gravity in $(d+1)$-dimensional de Sitter space $dS_{d+1}$ to a non-gravitational system in a certain lower dimensional space(time).

Due to lack of a stringy origin of de Sitter gravity, testing holography for any candidate of the dual theory could be a rather difficult task. A crucial step \cite{tb} in that direction is to estimate the dimension of the Hilbert space, which ought to give a finite number consistent with the Bekenstein-Hawking entropy if the candidate ``boundary'' theory is acceptable. Yet, one has another, perhaps even less conclusive way to start with, namely to compare spacetime symmetries of the bulk gravity with those in the proposed dual theories. Anyway, in many situations, symmetry and the number of degrees of freedom are the two things that may not sensitively depend on the details of the underlying dynamics. In the present paper we would like to make a couple of observations about spacetime symmetries in some possible holographic descriptions of de Sitter gravity, though occasionally we will also get in touch with counting dimensions of Hilbert spaces.

As is well-known, $(d+1)$-dimensional anti de Sitter space has the isometry group $SO(d,2)$, which is also the conformal group on the boundary of $AdS_{d+1}$ at infinity. From the symmetry point of view, it is thus quite natural to construct a CFT in $d$-dimensions as the holographic dual\footnote{This symmetry argument holds even if $SO(d,2)$ is partially broken by taking the quotient \cite{quotient} $AdS_{d+1}/\Gamma$, where $\Gamma$ is some discrete subgroup of $SO(d,2)$. In such a case both the bulk and the boundary symmetries reduce to the normalizer of $\Gamma$ in $SO(d,2)$.} to gravity in $AdS_{d+1}$ \cite{Mal}. Similarly, if the correct dual description of de Sitter gravity follows this pattern of holography, one expects that its spacetime symmetries will form the conformal group of a lower dimensional Euclidean space, suggesting that there may exist a $dS_{d+1}/CFT_{d}$ correspondence \cite{Stro}. In this correspondence the bulk symmetries agree perfectly with the boundary ones.

If, on the other hand, the correct realization of holography in de Sitter space patterns upon what one has seen in the BFSS model of M-theory, then the symmetry analysis will become a bit more involved. Recall that in flat background, the isometries of 11D supergravity form the Poincar\'e group $ISO(10,1)=SO(10,1)\ltimes{\mathbb R}^{10,1}$, which is apparently {\it not} the spacetime symmetries of any known boundary theory in $d=9+1$. In order to see holography, therefore, one has to work in the infinite momentum frame (IMF) \cite{BFSS}, or to perform discrete light cone quantization (DLCQ) \cite{DLCQ}. Both lead to a reduction of spacetime symmetries, and the reduced symmetries coincide with the (target) spacetime symmetries in Matrix Quantum Mechanics, which is the holographic dual to M-theory in flat background. Thus, if de Sitter gravity has a holographic description following this matrix-model pattern, we expect that its spacetime symmetries will reduce to a group $G_{\rm IMF}$ coming from taking a kind of ``IMF'', or to a group $G_{\rm DLCQ}$ (or its cousins) that corresponds to a ``light-cone'' subalgebra of the de Sitter algebra. 

The purpose of this paper is to investigate possible spacetime symmetries in the matrix description of de Sitter gravity, and to propose a concrete model based on one of such symmetry groups. We begin in Section 2 with a somewhat detailed description of $G_{\rm IMF}$ and $G_{\rm DLCQ}$ in $dS_{d+1}$, following algebraic considerations generalized from the flat space case. The former group, $G_{\rm IMF}$, is defined by the nonrelativistic limit of $SO(d,1)\subset SO(d+1,1)$. This is known as the Newton-Hooke group, a curved space deformation of Galilei symmetries. The latter group, $G_{\rm DLCQ}$, arises from a simple extension of the light-cone construction in flat space. We find that $G_{\rm DLCQ}\neq G_{\rm IMF}$ in de Sitter space, and discuss some possible physical implications of this result. In particular, it will be argued that matrix models with the symmetry group $G_{\rm IMF}$ look more plausible.

In Section 3 we study the one-particle quantum mechanical system that has $G_{\rm IMF}$ as its spacetime symmetries, and discuss a matrix model generalization of this system. As we shall see, the system will become quite complicated when there is a second central extension of the Newton-Hooke group. Such complication arises only if the bulk spacetime is 4-dimensional. For gravity in $dS_{4}$, our matrix model naturally incorporates issues such as holography and fuzziness, and, in certain scaling limit, we find a possible phase transition that may separate de Sitter gravity into different phases. The lower temperature phase, which corresponds to gravity in the perturbative regime, has a Hilbert space of infinite dimension. In the higher temperature phase where the perturbation theory breaks down, the dimension of the Hilbert space may become finite.

Finally, we summarize our conclusions in Section \ref{s4}.
\section{$G_{\rm IMF}$ versus $G_{\rm DLCQ}$}\label{s2}
For flat background the groups $G_{\rm IMF}$ and $G_{\rm DLCQ}$ are the same thing. We begin with a review of this fact and discuss its violation in de Sitter space. Such a violation forces us to determine whether $G_{\rm IMF}$, $G_{\rm DLCQ}$ are equally good for constructing a matrix model of de Sitter gravity, or one is better than the other. We will argue that it seems more reasonable to choose $G_{\rm IMF}$, rather than $G_{\rm DLCQ}$, as the ``boundary'' spacetime symmetries. More detailed descriptions of these groups will be given in Section \ref{ss2.1}--\ref{ss2.2} where, among other things, we can see explicitly that $G_{\rm IMF}\neq G_{\rm DLCQ}$. 

In the IMF approach to M-theory around flat space \cite{BFSS}, one compactifies 11D on a tiny spacelike circle, so that the spacetime symmetries reduce from $ISO(10,1)$ to $ISO(9,1)$. Performing a large Lorentz boost along this spacelike direction, we obtain a theory of IIA strings at weak coupling and in a sector of large D0-brane charges, which is effectively described by open string ground states that decouple from oscillator modes as well as from gravity. The fundamental degrees of freedom of this effective theory consists of heavy D0 branes, so the nonrelativistic limit (or, mathematically, the In\"on\"u-Wigner contraction) $ISO(9,1)\rightarrow G_{\rm IMF}$ can be trusted, where $G_{\rm IMF}$ denotes the Galilei group $Gal(9,1)$ in 9+1 dimensions. If one performs infinite boosts, the masses of D0 branes tend to infinity and we get a matrix theory with $G_{\rm IMF}$ as the exact target spacetime symmetries. This provides a holographic description of M-theory in the IMF after decompactifying the spacelike circle. 

Alternatively, we may apply DLCQ directly to 11D M-theory so that the spacetime symmetries are reduced to the group $G_{\rm DLCQ}$, which arises from the light-cone subalgebra of the full Poincar\'e symmetries $ISO(10,1)$. This again results in the Galilei group in 9+1 dimensions, {\it i.e.} $G_{\rm DLCQ}=Gal(9,1)$. The effective theory henceforth describes M-theory in a sector with fixed longitudinal momentum \cite{DLCQ}. That $G_{\rm DLCQ}$ is isomorphic to $G_{\rm IMF}$ is not accidental at all: According to Seiberg (and also Sen) \cite{S&S}, it is possible to perform a very large boost in the compactified M-theory while at the same time shrink the spacelike circle to a point, and, under the large boost limit, the shrunk spacelike circle will be Lorentz rotated to a light-like circle of finite size. This explains why DLCQ is equivalent to the IMF approach when formulating M(atrix) theory in flat space\footnote{As was shown by Hellerman and Polchinski \cite{HP}, quantum field theories compactified on a nearly lightlike circle in general suffer from strong coupling singularities if we actually take the infinite boost limit. This problem of divergences can be resolved by adding the stringy corrections from wrapping modes of branes \cite{bilal}.}. Evidently, the equivalence of the IMF and DLCQ reflects the fact that M-theory in flat background is invariant under Lorentz boosts.

As a warmup (and also for the purpose of fixing notations), let us recall a simple algebraic derivation of the fact $G_{\rm IMF}=G_{\rm DLCQ}$ in $(d+1)$-dimensional flat spacetime. The definition of $G_{\rm IMF}$ needs us to choose a subgroup $ISO(d-1,1)\subset ISO(d,1)$ and take the nonrelativistic limit. So let $P^{\mu}$, $J^{\mu\nu}$ ($0\leq\mu,\nu\leq d-1$) be the standard generators of $ISO(d-1,1)$, obeying the Poincar\'e algebra
\begin{eqnarray}
[P^{\mu},P^{\nu}]&=&0, \cr
[J^{\mu\nu},P^{\rho}]&=& i(\eta^{\mu\rho}P^{\nu}-\eta^{\nu\rho}P^{\mu}), \cr
[J^{\mu\nu},J^{\rho\sigma}]&=& i(\eta^{\mu\rho}J^{\nu\sigma}+\eta^{\nu\sigma}J^{\mu\rho}-\eta^{\mu\sigma}J^{\nu\rho}-\eta^{\nu\rho}J^{\mu\sigma}).
\label{Poincare}
\end{eqnarray}
here the signature of the Minkowski metric $\eta_{\mu\nu}$ is specified by $(-+\cdots+)$. We may introduce two parameters, $c$ (the speed of light) and $\mu$ (the mass), and write 
\begin{equation}
P^{0}=-P_{0}=-(\mu c\cdot{\bf 1}+\frac{1}{c}H), \quad
J^{0i}=cK^{i}\quad (i=1,\cdots,d-1).
\label{nonrel}
\end{equation}
Then substituting (\ref{nonrel}) into (\ref{Poincare}) and taking the nonrelativistic limit $c\rightarrow\infty$, we obtain the following algebra, which by definition is the algebra of $G_{\rm IMF}$:
\begin{eqnarray}
[P^{i},P^{j}]&=&[P^{i},H]\;=\;0,\cr
[J^{ij},P^{k}]&=&i(\delta^{ik}P^{j}-\delta^{jk}P^{i}),\quad [J^{ij},H]\;=\;0,\cr
[J^{ij},J^{mn}]&=&i(\delta^{im}J^{jn}+\delta^{jn}J^{im}-\delta^{in}J^{jm}-\delta^{jm}J^{in}),\cr
[P^{i},K^{j}]&=&-i\mu\delta^{ij}\cdot{\bf 1},\quad [H,K^{i}]=-iP^{i},\cr
[J^{ij},K^{k}]&=&i(\delta^{ik}K^{j}-\delta^{jk}K^{i}),\quad [K^{i},K^{j}]\;=\;0.
\label{Gal}
\end{eqnarray}
It is easy to see that this is the Galilei algebra $Gal(d-1,1)$ acting in $d$-dimensional spacetime, where $J_{ij}$ (the angular momentum) generates space rotations, $P_{i}$ (the momentum) and $H$ (the Hamiltonian) generate translations in space and in time, and $K^{i}$ is the generator of the Galilei boosts. The mass term $\mu\cdot{\bf 1}$ commutes with everything and thus it corresponds to a central extension. 

On the other hand, $G_{\rm DLCQ}$ is defined by the light-cone subalgebra\footnote{Here and below we (ab)use the same symbol for both a group and its corresponding Lie algebra.} of the Poincar\'e algebra in $(d+1)$-dimensions. Let us use $x^{a}$ with $0\leq a\leq d$ to describe a point of $\mathbb{R}^{1,d}$, and introduce the light-cone coordinates
\begin{equation}
x^{\pm}=\frac{1}{\sqrt{2}}(x^{0}\pm x^{d})
\end{equation}
together with the transverse ones $x^{i}$, $i=1,\cdots d-1$. In this frame, $\eta_{ij}=\delta_{ij}$, $\eta_{+-}=\eta_{-+}=-1$. The generators of the Poincar\'e algebra $ISO(d,1)$ along the light-cone and the longitudinal directions read
\begin{equation}
P^{\pm}=\frac{1}{\sqrt{2}}(P^{0}\pm P^{d}),\quad J^{\pm a}=\frac{1}{\sqrt{2}}(J^{0a}\pm J^{da}).
\end{equation}
Now if we define $K^{i}\equiv J^{+i}$, it is an easy matter to check that the set $\{P^{i},P^{\pm},K^{i},J^{ij}\}$ forms a closed subalgebra $G_{\rm DLCQ}$ of $ISO(d,1)$:
\begin{eqnarray}
[P^{i},P^{j}]&=&[P^{i},P^{\pm}]\;=\;[P^{+},P^{-}]\;=\;0,\cr
[J^{ij},P^{k}]&=&i(\delta^{ik}P^{j}-\delta^{jk}P^{i}),\quad [J^{ij},P^{\pm}]\;=\;0,\cr
[J^{ij},J^{mn}]&=&i(\delta^{im}J^{jn}+\delta^{jn}J^{im}-\delta^{in}J^{jm}-\delta^{jm}J^{in}),\cr
[P^{i},K^{j}]&=&i\delta^{ij}P^{+},\quad [P^{+},K^{i}]\;=\;0,\quad [P^{-},K^{i}]=iP^{i},\cr
[J^{ij},K^{k}]&=&i(\delta^{ik}K^{j}-\delta^{jk}K^{i}),\quad [K^{i},K^{j}]\;=\;0.
\label{Gal+}
\end{eqnarray}
For a comparison of (\ref{Gal+}) and (\ref{Gal}), note that $P_{a}=\eta_{ab}P^{b}$ generate translations along $x^{a}$. So $P_{+}=-P^{-}=-\frac{1}{\sqrt{2}}(P^{0}-P^{d})=\frac{1}{\sqrt{2}}(P_{0}+P_{d})$  shifts the light-cone time $x^{+}\equiv t$ by a unit, thus representing the light-cone Hamiltonian $H$. On the other hand, $P_{-}=-P^{+}=\frac{1}{\sqrt{2}}(P_{0}-P_{d})$ generates a boost along the longitudinal direction $x^{-}$, which commutes with everything in the subalgebra and hence defines a central element $P_{-}\equiv \mu\cdot{\bf 1}$. Now, with the substitution $P^{+}\rightarrow -\mu\cdot{\bf 1}$, $P^{-}\rightarrow -H$, we see that (\ref{Gal+}) becomes exactly (\ref{Gal}). Thus we get the desired relation $G_{\rm DLCQ}=G_{\rm IMF}$ in flat space. 

Notice that the spectrum $M^{2}=-P_{a}P^{a}=2\mu\cdot H-(P^{i})^{2}$ of a single relativistic particle in $(d+1)$-dimensions gives
\begin{equation}
H=\frac{P_{i}^{2}}{2\mu}+U,\quad U\equiv M^{2}/(2\mu)
\label{Hamiltonian}
\end{equation}
which governs a nonrelativistic dynamical system in $d$-dimensions, with $U$, $\mu$ playing the roles of internal energy and mass, respectively. It is possible to extend this system using matrix degrees of freedom and incorporating Galilei/SUSY invariant interaction terms. In order for such a ``boundary'' theory to have maximal supercharges, the dimension $d$ and the interaction terms cannot be completely arbitrary; the resulting theory is just the BFSS matrix model for $d=10$ (or its T-dual cousins for other $d$), which is holographically equivalent to the original supergravity in 11D (or other versions via compactifications).

When we try to extend the above analysis to the de Sitter case, some problems will arise. Of course, the main difficulty is that de Sitter gravity has not yet been successfully embedded into string/M-theory\footnote{Though there are suggestions that de Sitter geometry can arise as classical solutions of some nonstandard theories, or from some nonstandard compactification procedures; see e.g. \cite{GKK}\cite{MM} for a few recent references.}, and it is difficult to detect, even at the kinetic level, its properties. All we know kinetically is that spacetime symmetries of the underlying theory in $dS_{d+1}$ (if it exists) should form the de Sitter group $SO(d+1,1)$. It is not clear at all, for example, whether we can construct some simple objects (similar to D0 branes) in the compactification\footnote{The compactification $dS_{d+1}\rightarrow dS_{d}$ could be described by, e.g., sending one spacelike coordinate in the hyperboloid equation $-(y^{0})^{2}+(y^{1})^{2}+\cdots+(y^{d+1})^{2}=R^{2}$ to zero.} of $dS_{d+1}$ to $dS_{d}$, so that they get decoupled from all other degrees of freedom when the infinite mass limit is taken. The existence of such objects would be essential in an attempt to derive a matrix quantum mechanics with the exact spacetime symmetry group $G_{\rm IMF}$, which is now defined by the nonrelativistic limit of $SO(d,1)\subset SO(d+1,1)$. 

Also, we know pretty little about DLCQ in $dS_{d+1}$ at present; hence it is quite difficult to address questions such as whether a matrix model based on DLCQ is sensible for describing de Sitter gravity. As we shall see explicitly, if the ``light-cone subgroup'' $G_{\rm DLCQ}$ of $SO(d+1,1)$ is formally defined in $dS_{d+1}$ by extending the flat space construction, we will end up with, comparing to the flat space result, a somewhat negative output: $G_{\rm DLCQ}\neq G_{\rm IMF}$. Physically this should not be very surprising, since while the underlying theory in $dS_{d+1}$ is still expected to be Lorentz boost invariant, we have no clear stringy reasoning to guarantee that either the nonrelativistic limit of $SO(d,1)$ or taking light-cone like subalgebras of $SO(d+1,1)$ will correspond to a certain decoupling limit of the theory, let along a reasoning to guarantee the equivalence of the IMF and DLCQ. Thus, even from the kinetic point of view, it seems hard to find a correct matrix description of de Sitter gravity. 

The violation of $G_{\rm IMF}=G_{\rm DLCQ}$ in de Sitter space can be seen most easily from a purely algebraic argument. The algebra $G_{\rm IMF}$ in fact contains more generators than $G_{\rm DLCQ}$ does. To see this, first note that the number of generators in $G_{\rm IMF}$ should be the same as those in $SO(d,1)$, since taking the nonrelativistic contraction does not annihilate Lie algebra elements. Thus, apart from possible central elements, we have
\begin{equation}
\dim G_{\rm IMF}=\dim SO(d,1)=\frac{d(d+1)}{2}.
\label{dimIMF}
\end{equation}
Next we consider generators in $G_{\rm DLCQ}$. Recall that the de Sitter algebra $SO(d+1,1)$ takes the form
\begin{eqnarray}
[P^{a},P^{b}]&=&\frac{i}{R^{2}}J^{ab}, \quad
[J^{ab},P^{c}]\;=\; i(\eta^{ac}P^{b}-\eta^{bc}P^{a}), \cr
[J^{ab},J^{cd}]&=& i(\eta^{ac}J^{bd}+\eta^{bd}J^{ac}-\eta^{ad}J^{bc}-\eta^{bc}J^{ad}),
\label{deSitter+}
\end{eqnarray}
which resembles the Poincar\'e algebra in $(d+1)$-dimensions but now the momentum operators $P_{a}$ are no longer commuting. If we define the light-cone objects $P^{\pm}=\frac{1}{\sqrt{2}}(P^{0}\pm P^{d})$, $K^{i}=J^{+i}$ naively as in the flat space case, the generators $\{P^{i},P^{\pm},K^{i},J^{ij}\}$ do not constitute a closed subalgebra, nor is the operator $P^{+}$ a center. (To be a little more concrete: the commutator $[P^{i},P^{-}]$ is proportional to $J^{-i}$, which lives outside the linear space spanned by $P^{i}$, $P^{\pm}$, $K^{i}$ and $J^{ij}$; also, we have $[P^{+},P^{-}]\propto J^{+-}\neq 0$.) This forces us to consider $G_{\rm DLCQ}$ as a subset of $\{P^{i},P^{\pm},K^{i},J^{ij}\}$ which, on the one hand, should form a closed algebra and, on the other hand, should contain $P^{+}$ as a center\footnote{The requirement that $P^{+}$ is a center validates the application of DLCQ to a sector with fixed longitudinal momentum $P^{+}=-\mu$. Otherwise, $P^{+}$ will be a nontrivial operator and we cannot assign it to a number.}. Mathematically, such a $G_{\rm DLCQ}$ can be defined by the {\it centralizer} of $P^{+}$ in $SO(d+1,1)$. In Section \ref{ss2.2}, we will find that this centralizer is spanned by $\{P^{+},K^{i},J^{ij}\}$. Consequently, the number of generators in $G_{\rm DLCQ}$ (not including the center $P^{+}$) reads
\begin{equation}
\dim G_{\rm DLCQ} = \dim [SO(d-1)\ltimes\mathbb{R}^{d-1}]=\frac{d(d-1)}{2},
\label{dimDLCQ}
\end{equation}
which is indeed less than $\dim G_{\rm IMF}$.

The above result, namely $G_{\rm DLCQ}\neq G_{\rm IMF}$, may be interpreted as a kind of kinetic no-go. It gives an algebraic obstruction in extending the boost argument \cite{S&S} to the de Sitter case. We have two kinetically inequivalent ways to formulate matrix models, one based on $G_{\rm IMF}$ and the other based on $G_{\rm DLCQ}$, and it is not easy to see which one is acceptable. Moreover, if one such model is the correct description of gravity in $dS_{d+1}$, we may ask why the other would not, given that Lorentz symmetries constitute a subgroup of $SO(d+1,1)$? The answer to this question would be closely related to the observation \cite{HP} that in flat space, when one passes from IMF to DLCQ, the longitudinal zero modes in a $d$-dimensional QFT behave effectively as some $(d-1)$-dimensional strongly coupled theory. In flat background the strong coupling singularities can be compensated by wrapped D-branes \cite{bilal}, but due to lack of stringy origins of de Sitter geometry, such a simple compensation is unlikely to occur here. Thus, from the dynamical point of view, it might not be too odd to see that the number in Eq.(\ref{dimIMF}) is greater than (\ref{dimDLCQ}) or, in a more enlightening way, that $G_{\rm IMF}$ acts in $d$-dimensions while $G_{\rm DLCQ}$ looks like a symmetry group acting in $(d-1)$-dimensions. Although an actual dynamical analysis would be difficult to perform, we expect that in order to formulate a ``boundary'' theory with $G_{\rm DLCQ}$ as the spacetime symmetries, one has to resolve the problem of strong coupling singularities from the outset. This suggests that $G_{\rm DLCQ}$ may not be the correct symmetry group to start with, and $G_{\rm IMF}$ should be more plausible for constructing a matrix model of de Sitter gravity.
\subsection{$G_{\rm IMF}$: Newton-Hooke from $dS_{d+1}$}\label{ss2.1}
We have seen that spacetime symmetries in the holographic description of M-theory \cite{BFSS} form the Galilei group $Gal(9,1)$. As mentioned, this group comes from a two-step construction: (i) picking up a subgroup $ISO(9,1)$ of the full Poincar\'e group $ISO(10,1)$ in eleven dimensions, and (ii) taking the In\"on\"u-Wigner contraction of this subgroup under the nonrelativistic limit. Physically, step (i) is achieved by compactifying 11D M-theory on a small spacelike circle, and thus we have weakly coupled IIA strings with KK excitations. Step (ii) corresponds to performing infinite Lorentz boost along the spacelike circle, so all the KK excitations become infinitely heavy and hence decouple from other degrees of freedom. 

In our de Sitter case, therefore, it is also tempting to construct a ``boundary'' symmetry group $G_{\rm IMF}$ by taking the nonrelativistic contraction of the subgroup $SO(d,1)\subset SO(d+1,1)$. This relies mainly on a mathematical analogue of the flat space construction rather than on a solid physical basis. From the physical point of view, the first step to get such a $G_{\rm IMF}$ would be the compactification of de Sitter gravity in $dS_{d+1}$ to $dS_{d}$. It is quite easy to see that $dS_{d}$ can be embedded as a hypersurface in $dS_{d+1}$. For example, in the global parametrization of $dS_{d+1}\cong\mathbb{R}\times S^{d}$, the induced metric takes the form
\begin{equation}
ds^{2}=-dt^{2}+R^{2}\cosh^{2}\left(\frac{t}{R}\right)d\Omega_{d}^{2}
\label{Gmetric}
\end{equation}
So if we write the standard metric on the unit sphere $S^{d}$ as $d\Omega_{d}^{2}=d\chi^{2}+\sin^{2}\chi\cdot d\Omega_{d-1}^{2}$, we find that the hypersurface at $\chi=\frac{\pi}{2}$ gives precisely an embedding $dS_{d}\hookrightarrow dS_{d+1}$. This embedding is somewhat warped and we may work with certain generalized dimensional reduction procedure. Since $\chi$ is a compact coordinate, we do not have to consider localization properties of bulk fields as in the AdS case. 

After compactifying to $dS_{d}$, we obtain a theory with the spacetime symmetries $SO(d,1)$, which is again the de Sitter group (but in $d$-dimensional spacetime). So our second step is to consider the sector of heavy KK modes, hoping that it will eventually decouple when the limit of infinite masses is taken. The spacetime symmetries in this sector are then described by the nonrelativistic version of $SO(d,1)$, and this defines $G_{\rm IMF}$ in de Sitter space. Mathematically, this group arises from an In\"on\"u-Wigner contraction of $SO(d,1)$ \cite{BL}, which is known as the Newton-Hooke group. We will therefore use the symbol $NH(d-1,1)$ to denote such a group acting in $d$-dimensional ``boundary'' spacetime. 

An explicit description of $NH(d-1,1)$ is given as follows. Let us start with the de Sitter algebra $SO(d,1)$. Choose a set of hermitian generators $P^{\mu}$, $J^{\mu\nu}$ with $\mu,\nu=0,1,\cdots,d-1$, such that their commutation relations take the form of (\ref{deSitter+}) but in lower-dimensions:
\begin{eqnarray}
[P^{\mu},P^{\nu}]&=&\frac{i}{R^{2}}J^{\mu\nu}, \quad
[J^{\mu\nu},P^{\rho}]\;=\; i(\eta^{\mu\rho}P^{\nu}-\eta^{\nu\rho}P^{\mu}), \cr
[J^{\mu\nu},J^{\rho\sigma}]&=& i(\eta^{\mu\rho}J^{\nu\sigma}+\eta^{\nu\sigma}J^{\mu\rho}-\eta^{\mu\sigma}J^{\nu\rho}-\eta^{\nu\rho}J^{\mu\sigma}).
\label{deSitter}
\end{eqnarray}
Clearly, under the flat space limit $R\rightarrow\infty$, (\ref{deSitter}) contracts to the algebraic relations defining $ISO(d-1,1)$. This is the usual way to make  In\"on\"u-Wigner contraction of the de Sitter group. There is a second way to perform contraction, which is achieved by taking the nonrelativistic limit of (\ref{deSitter}) while keeping the (rescaled) spacetime curvature finite \cite{BL}. Hence we make the substitutions similar to (\ref{nonrel})
\begin{equation}
P^{0}\rightarrow -(\mu c\cdot{\bf 1}+\frac{1}{c}H),\quad P^{i}\rightarrow P^{i},\quad J^{0i}\rightarrow cK^{i},\quad J^{ij}\rightarrow J^{ij},\quad R\rightarrow cR
\label{nonreldS}
\end{equation}
which gives, after taking the nonrelativistic limit $c\rightarrow\infty$, the Newton-Hooke algebra:
\begin{eqnarray}
[P^{i},P^{j}]&=&0,\quad [H,P^{i}]\;=\;-\frac{i}{R^{2}}K^{i}, \cr
[J^{ij},P^{k}]&=& i(\delta^{ik}P^{j}-\delta^{jk}P^{i}), \quad [J^{ij},H]\;=\;0,\cr
[J^{ij},J^{mn}]&=& i(\delta^{im}J^{jn}+\delta^{jn}J^{im}-\delta^{in}J^{jm}-\delta^{jm}J^{in}),\cr
[P^{i},K^{j}]&=&-i\mu\,\delta^{ij}\cdot{\bf 1},\quad [H,K^{i}]\;=\;-iP^{i},\cr
[J^{ij},K^{k}]&=&i(\delta^{ik}K^{j}-\delta^{jk}K^{i}),\quad [K^{i},K^{j}]\;=\;0.
\label{NH}
\end{eqnarray}
Evidently, this will contract further to the Galilei algebra (\ref{Gal}) in the flat space limit $R\rightarrow\infty$, in which the Hamiltonian becomes invariant under spatial translations. The number of non-central generators in (\ref{NH}) is thus $\dim G_{\rm IMF}=2d-1+(d-1)(d-2)/2=d(d+1)/2$.

It is not difficult to work out the spacetime transformations generated by this algebra:
\begin{eqnarray}
x^{i}&\rightarrow & x'^{i}={\cal R}{^{i}}_{j}\cdot x^{j}+v^{i}R\,\sinh\frac{t}{R}+a^{i}\,\cosh\frac{t}{R}\cr
t & \rightarrow & t'=t+b
\label{NHtran}
\end{eqnarray}
where ${\cal R}=({\cal R}{^{i}}_{j})\in SO(d-1)$ is a space rotation generated by the angular momentum $J^{ij}$, $v^{i}$ is a ``velocity'' corresponding to the boost $K^{i}$, and $a^{i}$, $b$ are spacetime translations generated by the momentum $P_{i}$ and the Hamiltonian $H$, respectively. The Lie algebra generators of such transformations can be described by the tangent vectors to the group orbit at identity. Thus, for example, the generators of spatial translations are represented by the vector fields
\begin{equation}
P_{i}\equiv -i\left.\frac{\partial x'^{j}}{\partial a^{i}}\right|_{a=b=v=0,\,{\cal R}=1}\cdot\frac{\partial}{\partial x^{j}}=-i\cosh\frac{t}{R}\cdot\frac{\partial}{\partial x^{i}}.
\end{equation}
In a similar way, we can write down the time translation generator $H$, the  boosts $K_{i}$ as well as the angular momentum operators $J_{ij}$ in terms of the following vector fields:
\begin{equation}
H=-i\frac{\partial}{\partial t},\quad K_{i}=-iR\,\sinh\frac{t}{R}\cdot\frac{\partial}{\partial x^{i}},\quad
J_{ij}=-i\left(x_{i}\frac{\partial}{\partial x^{j}}-x_{j}\frac{\partial}{\partial x^{i}}\right).
\end{equation}
It is straightforward to check that these operators obey the commutation relations (\ref{NH}), up to the central extension by the mass parameter $\mu$. Actually, central terms cannot be obtained in this simple representation; they are related to a projective representation of the generators $P_{i},K_{i}$ acting on some Hilbert space \cite{BL} \cite{AHPS}.

Let us give some more discussions on central extensions. In general dimensions, the mass operator $\mu\cdot{\bf 1}$ is the only possible central terms of the Newton-Hooke algebra $NH(d-1,1)$. Thus, if $\mu\neq 0$, the commutation relations $[P^{i},K^{j}]=-i\mu\delta^{ij}$ and $[K^{i},K^{j}]=0$ suggest that in quantum mechanics, the boost operator $K^{i}$ is essentially related to the position operator $X^{i}$ through $K^{i}\sim\mu X^{i}$. Accordingly, the generator $H$ in (\ref{NH}) can be realized by the following single particle Hamiltonian
\begin{equation}
H=\frac{1}{2\mu}\left(P_{i}^{2}-\frac{1}{R^{2}}K_{i}^{2}\right)\sim \frac{1}{2\mu}P_{i}^{2}-\frac{\mu}{2R^{2}}X_{i}^{2},
\label{Halg}
\end{equation}
which contains a potential of inverted harmonic oscillators. This realization shows that the center $\mu$ can indeed be interpreted as a mass parameter.

Now we specify to $d=2+1$. This special case is important since the corresponding bulk space is $dS_{4}$, {\it i.e.}, 4-dimensional de Sitter space. On the ``boundary'', we have the symmetries $NH(2,1)$ and, by rewriting $J^{ij}=\epsilon^{ij}J$ (with $i=1,2$), we see that the algebra (\ref{NH}) reduces to
\begin{eqnarray}
[P^{i},P^{j}]&=&0,\quad [H,P^{i}]\;=\;-\frac{i}{R^{2}}K^{i}, \cr
[J,P^{i}]&=& i\epsilon^{ij}P_{j}, \quad [J,H]\;=\;0,\cr
[P^{i},K^{j}]&=&-i\mu\,\delta^{ij}\cdot{\bf 1},\quad [H,K^{i}]\;=\;-iP^{i},\cr
[J,K^{i}]&=&i\epsilon^{ij}K_{j},\quad [K^{i},K^{j}]\;=\;0.
\label{NH2+1}
\end{eqnarray}
Since the rotation subgroup of $NH(2,1)$ is simply the abelian group $SO(2)$ generated by $J$, the Newton-Hooke algebra in $2+1$-dimensions allows some ``exotic'' central extensions \cite{AHPS}. Thus, in addition to the central element $\mu$ which is interpreted as a mass parameter, we can introduce a second parameter $\kappa$ so that the boost generators $K^{i}$ are no longer commuting: $[K^{i},K^{j}]=i\kappa\epsilon^{ij}{\bf 1}$. This will force the commutator between $P^{i}$ and $P^{j}$ nonvanishing. In fact, according to the Jacobian identity $[[H,P^{i}],K^{j}]+[[P^{i},K^{j}],H]+[[K^{j},H],P^{i}]=0$, one must have $[P^{i},P^{j}]=-i\frac{\kappa}{R^{2}}\epsilon^{ij}{\bf 1}$. Also, it is possible to introduce a third central element $\lambda$ so that $J$ and $H$ no longer commute with each other: $[J,H]=i\lambda{\bf 1}$. Summarizing the above yields:
\begin{eqnarray}
[P^{i},P^{j}]&=&-i\frac{\kappa}{R^{2}}\epsilon^{ij}{\bf 1},\quad [H,P^{i}]\;=\;-i\frac{1}{R^{2}}K^{i}, \cr
[J,P^{i}]&=& i\epsilon^{ij}P_{j}, \quad [J,H]\;=\;i\lambda{\bf 1},\cr
[P^{i},K^{j}]&=&-i\mu\,\delta^{ij}{\bf 1},\quad [H,K^{i}]\;=\;-iP^{i},\cr
[J,K^{i}]&=&i\epsilon^{ij}K_{j},\quad [K^{i},K^{j}]\;=\;i\kappa\epsilon^{ij}{\bf 1}.
\label{NH2+1'}
\end{eqnarray}
We thus get the maximally centrally-extended Newton-Hooke algebra in (2+1)-dimensions \cite{AMO}, which is similar to the familiar flat space case where the Galilei algebra $Gal(2,1)$ admits a 3-parameter central extension \cite{LL}. 

Note that just as in the Galilei case \cite{BGK} \cite{LSZ}, the third center $\lambda$ is physically uninteresting because: (i) its presence does not allow us to integrate the Lie algebra to a Lie group, (ii) it is not obtainable by taking the nonrelativistic limit of $SO(3,1)\subset SO(4,1)$, hence not relevant to the bulk physics, and (iii) unlike (\ref{Halg}), there is no natural realization of the Hamiltonian $H$ in terms of the operators $P^{i}$ and $K^{i}$, if $\lambda\neq 0$. For these reasons, we will simply discard the third central extension and always set $\lambda=0$.

The second center, $\kappa$, can be obtained from the nonrelativistic contraction of $SO(3,1)$ if instead of $J^{ij}\rightarrow J^{ij}$ in (\ref{nonreldS}), we make the substitution $J^{ij}\rightarrow J^{ij}-c^{2}\kappa\epsilon^{ij}{\bf 1}$ before taking the limit $c\rightarrow\infty$. This suggests that similar to the mass parameter, $\kappa$ would also correspond to some charge (or other quantum number) arising in the KK reduction $dS_{4}\rightarrow dS_{3}$ of the underlying theory. Moreover, we can realize the Hamiltonian $H$ in the algebra (\ref{NH2+1'}) in terms of $P_{i}$, $K_{i}$, $\mu$ and $\kappa$, by
\begin{equation}
H=\frac{1}{2\mu[1+(\frac{\kappa}{\mu R})^{2}]}
\left(P_{i}^{2}-\frac{1}{R^{2}}K_{i}^{2}+\frac{2\kappa}{\mu R^{2}}\,\epsilon_{ij}P^{i}K^{j}\right).
\label{Halg-k}
\end{equation}
We see that this realization differs from (\ref{Halg}) by a change in the mass $\mu\rightarrow \mu[1+(\frac{\kappa}{\mu R})^{2}]$ plus an additional term proportional to $\epsilon_{ij}P^{i}K^{j}$. 

The additional term in (\ref{Halg-k}) may raise some subtleties in the construction of a dynamical model. At first sight, this term would give us a Chern-Simons like coupling $\epsilon_{ij}\dot{X}^{i}X^{j}$ with the noncommutative position operators $X^{i}$ obeying $[X^{i},X^{j}]\sim i\frac{\kappa}{\mu^{2}}\epsilon^{ij}$. Such a coupling could be enhanced at $\mu\approx 0$. Thus, one might expect that the matrix Chern-Simons model proposed in \cite{ML} would correspond to a limiting case of the centrally-extended Newton-Hooke symmetries. But just as in the flat space case \cite{JN}, it is possible to invoke a version of the Foldy-Wouthuysen transformation to find some canonical pairs $x^{i},p_{i}$ with $[x^{i},p_{j}]=i\delta^{i}_{j}$, $[x^{i},x^{j}]=[p_{i},p_{j}]=0$, so that the algebra generators $K_{i}$ and $P_{i}$ are realized by
\begin{eqnarray}
K_{i}&=&\mu^{*} x_{i}-\frac{\kappa}{2\mu^{*}}\,\epsilon_{ij}\,p^{j}\cr
P_{i}&=&p_{i}-\frac{\kappa}{2R^{2}}\,\epsilon_{ij}\,x^{j}
\label{canKP}
\end{eqnarray}
where $\mu^{*}$ is the effective mass determined by the algebraic relations $[P_{i},K_{j}]=-i\mu\delta_{ij}$:
\begin{equation}
\mu^{*}-\frac{\kappa^{2}}{4\mu^{*}R^{2}}=\mu\quad\Rightarrow\quad\mu^{*}=\frac{\mu+\sqrt{\mu^{2}+(\kappa/R)^{2}}}{2}.
\label{effMass}
\end{equation}
Using these canonical variables (and in particular the commutative position operators $x^{i}$), we can express the Hamiltonian (\ref{Halg-k}) in the form
\begin{equation}
H=\frac{1}{2\mu^{*}}\,p_{i}^{2}-\frac{\mu^{*}}{2R^{2}}\,x_{i}^{2}
\label{Halg-can}
\end{equation}
without Chern-Simons like couplings. This again gives rise to the pure upside-down harmonic oscillators, but now the mass becomes $\mu^{*}$. Note that this consideration is valid even if $\mu=0$. 

However, similar to the observation made in \cite{BGK} for the flat space case, one can see that the change of the operator basis $(K_{i},P_{i})\rightarrow (x_{i},p_{i})$ specified by (\ref{canKP}) is really a redefinition of physical observables rather than a canonical transformation in phase space. In particular, (\ref{canKP}) does not correspond to the Noether charges $K_{i}$, $P_{i}$ derived directly from a system with the Hamiltonian (\ref{Halg-can}). (For such a system we could expect $K_{i}\sim \mu^{*}x_{i}$ and $P_{i}\sim p_{i}=\mu^{*}\dot{x}_{i}$.) One thus suspects that adding some Chern-Simons like couplings is essential for realizing the Newton-Hooke symmetries in the presence of the second central extension $\kappa$. In Section \ref{s3}, we will provide a concrete model based on the doubly-extended Newton-Hooke group, whose Lagrangian contains both terms $\epsilon_{ij}x^{i}\dot{x}^{j}$ and $\epsilon_{ij}\dot{x}^{i}\ddot{x}^{j}$ of the Chern-Simons type.
\subsection{$G_{\rm DLCQ}$: Generalized Light-cone Gauges}\label{ss2.2}
In flat space $G_{\rm DLCQ}$ is defined by the subalgebra of $ISO(d,1)$ arising from fixing the light-cone gauge $P^{+}={\rm const}$. We have seen that this is isomorphic to the algebra $G_{\rm IMF}$ constructed from the nonrelativistic limit of $ISO(d-1,1)\subset ISO(d,1)$. Now in the de Sitter case, given the bulk symmetries ${\cal G}_{d+1}=SO(d+1,1)$, how to appropriately define its light-cone subalgebra $G_{\rm DLCQ}$? 

Our proposal, which is motivated by the foregoing discussions, can be described generally as follows. Choose $\Pi\in {\cal G}_{d+1}$ to be a linear combination of the momentum operators $P_{a}$,
\begin{equation}
\Pi=\sum_{a=0}^{d}\xi^{a}P_{a}
\label{center}
\end{equation}
where $\xi^{a}$ are fixed numerical coefficients. We want to construct a maximal subalgebra $G_{\rm DLCQ}$ of ${\cal G}_{d+1}$ such that (\ref{center}) is its central element. Mathematically this definition says that $G_{\rm DLCQ}$ is the {\it centralizer} of $\Pi$ in ${\cal G}_{d+1}$, namely 
\begin{equation}
G_{\rm DLCQ}=Z(\Pi)\equiv\left\{T\in {\cal G}_{d+1}\right|\left.[T,\Pi]=0\right\}.
\label{centralizer}
\end{equation}
This is somewhat similar but not identical to Wigner's little group. For comparison, recall that a little group is defined as an isotropy subgroup of $SO(3,1)$ that keeps a given Lorentz vector $p^{\mu}$ (outside $SO(3,1)$) invariant. Physically, the choice of (\ref{centralizer}) allows us to work in a sector of the fixed momentum $\Pi=\mu\cdot{\bf 1}$, which defines a generalized light-cone gauge.

The centralizer so defined enjoys several nice properties. First of all, $Z(\Pi)$ is a subalgebra of ${\cal G}_{d+1}$ since, if $T,T'\in Z(\Pi)$, then both a linear combination of $T$, $T'$ and the commutator between $T$, $T'$ commute with $\Pi$. Second, if $C(\Pi)$ is a Cartan subalgebra of ${\cal G}_{d+1}$ containing $\Pi$, then $C(\Pi)\subset Z(\Pi)$. In particular the dimension of $Z(\Pi)$ should not be less than the rank of ${\cal G}_{d+1}$,
\begin{equation}
\dim Z(\Pi)\geq {\rm rank}({\cal G}_{d+1}).
\label{dimZ}
\end{equation}
Similarly, if $\Pi$ is contained in two distinct Cartan subalgebras $C(\Pi)$ and $C'(\Pi)$ of ${\cal G}_{d+1}$, then we must have $C(\Pi)\cup C'(\Pi)\subset Z(\Pi)$ and in this case the inequality (\ref{dimZ}) becomes strict (the property clearly extends to the case of multi Cartan subalgebras that contain $\Pi$). Such $\Pi$ is called a singular element of ${\cal G}_{d+1}$. So if one takes $\Pi$ to be singular, the centralizer would not be too small.
Finally, if ${\cal G}_{d+1}$ has a Cartan subalgebra $Span\{H_{i}\}$ containing $\Pi$ and allows a root system decomposition in the form ${\cal G}_{d+1}=\oplus_{i}H_{i}\oplus_{\alpha\in\Delta_{+}}(E_{\alpha}\oplus E_{-\alpha})$, then any element of $Z(\Pi)$ can be expressed by a linear combination
\begin{equation}
A=\sum_{\alpha\in\Delta_{+},\;\alpha(\Pi)=0}(\lambda_{\alpha}E_{\alpha}+\lambda_{-\alpha}E_{-\alpha})+\sum_{i}\lambda_{i}H_{i}.
\end{equation}
Of course this property holds for any simple, compact Lie groups.

A practical way to find $Z(\Pi)$ for the de Sitter group ${\cal G}_{d+1}=SO(d+1,1)$ can be described by the following. Let us choose the basis $\{P_{a},J_{ab}\}$ of ${\cal G}_{d+1}$ as before, which satisfies the commutation relations (\ref{deSitter+}), and let us write $\Pi$ as a linear combination (\ref{center}) of the momentum operators. We expand each element of ${\cal G}_{d+1}$ in terms of $P_{a}$ and $J_{ab}$,
\begin{equation}
A=\sum_{a}\lambda^{a}P_{a}+\frac{1}{2}\sum_{ab}\omega^{ab}J_{ab},
\label{expansion}
\end{equation}
with $\omega_{ab}=-\omega_{ba}$. When $A\in Z(\Pi)$, the coefficients $\lambda^{a}$ and $\omega^{ab}$ should obey some constraints determined by $[A,\Pi]=0$. Now using (\ref{deSitter+}), it is not difficult to compute
\begin{equation}
[A,\Pi]=\frac{i}{2R^{2}}(\lambda^{a}\xi^{b}-\lambda^{b}\xi^{a})J_{ab}+i\omega^{ab}\xi_{a}P_{b},
\label{[A,Pi]}
\end{equation}
so that the condition of $A$ commuting with $\Pi$ gives rise to the constraints
\begin{equation}
\lambda^{a}\xi^{b}-\lambda^{b}\xi^{a}=0, \quad
\omega^{ab}\xi_{a}=0.
\label{constraint}
\end{equation}
Thus, for fixed $\xi^{a}$ (namely $\Pi$ is held fixed), substituting each solution $(\lambda^{a},\omega^{ab})$ of (\ref{constraint}) into (\ref{expansion}) yields an element $A$ of $Z(\Pi)$ and, conversely, each element of $Z(\Pi)$ corresponds to a solution of (\ref{constraint}).

Note that the above constraints are derived for a finite de Sitter size $R$. In the flat space limit $R\rightarrow\infty$, we can get only the second constraint in (\ref{constraint}), while the first constraint should not be imposed. This originates from the fact that when $R$ tends to infinity, the first term in the right hand side of (\ref{[A,Pi]}) does not appear. One consequence of this is that the subgroup $G_{\rm DLCQ}=Z(\Pi)$ of the Poincar\'e symmetries ${\cal G}_{d+1}=ISO(d,1)$ has more generators than $G_{\rm DLCQ}$ in the the de Sitter case, where ${\cal G}_{d+1}=SO(d+1,1)$. In particular, $Z(\Pi)$ contains all the momentum operators $P_{a}$ in the flat space limit.
\subsubsection{Returning to Flat Space}
Before working out $Z(\Pi)$ explicitly for the de Sitter group, let us first consider the case of ${\cal G}_{d+1}=ISO(d,1)$ in flat space, checking that the Galilei subalgebra (\ref{Gal+}) can indeed appear as the centralizer of the longitudinal momentum $P_{-}$. For this purpose, we take $\xi^{0}=1$, $\xi^{i}=0$ ($1\leq i\leq d-1$), and $\xi^{d}=-1$, so that (\ref{center}) gives the  $\Pi=P_{-}$. Substituting these data into the second condition in (\ref{constraint}), we find (notice that $\xi_{0}=-\xi^{0}=-1$ and $\xi_{d}=\xi^{d}=-1$):
\begin{equation}
(\lambda^{a},\omega^{ij})=\hbox{free parameters},\quad
\omega^{0i}+\omega^{di}=0.
\end{equation}
Accordingly, the elements of $Z(\Pi)$ all take the form
\begin{equation}
A=\lambda^{a}P_{a}+\frac{1}{2}\,\omega^{ij}J_{ij}+\omega^{0i}(J_{0i}-J_{di}).
\label{solutions}
\end{equation}
It follows that $Z(\Pi)$ is generated by $P_{i}$, $P_{\pm}$ (with $P_{-}=\Pi$ being a center), $J_{ij}$, and $J_{0i}-J_{di}=-(J{^{0}}_{i}+J{^{d}}_{i})=-\sqrt{2}\eta_{ij}K^{j}$. Hence $Z(\Pi)$ is exactly the Galilei subalgebra (\ref{Gal+}).

In the above derivation we take $\xi^{a}=(1,0,\cdots,0,-1)$, which is a special case of $\xi^{a}$ being null, {\it i.e.}, $\langle\xi,\xi\rangle=\eta_{ab}\,\xi^{a}\xi^{b}=0$. Actually, given any null vector $\xi^{a}$ in the above construction, we will arrive at the same Galilei subalgebra $Z(\Pi)$, perhaps with a different embedding into $ISO(d,1)$. The easiest way to look at this is to recall: (i) any null vector $\xi^{a}$ can be Lorentz rotated to the form $\xi'^{a}=\Lambda{^{a}}_{b}\,\xi^{b}\propto(1,0\cdots,0,-1)$, and (ii) redefining the algebra generators of $ISO(d,1)$ via the Lorentz rotations $P'^{a}=\Lambda{^{a}}_{b}P^{b}$, $J'^{ab}=\Lambda{^{a}}_{c}\Lambda{^{b}}_{d}J^{cd}$ preserves the Poincar\'e algebra. Thus, since $\Pi=\xi^{a}P_{a}=\xi'^{a}P'_{a}$, we find that $P'^{a}$, $K'^{i}$ and $J'^{ij}$ span the Galilei subalgebra $Z(\Pi)$, and these generators are linear combinations of the original Poincar\'e generators, not necessarily living in the same subspace spanned by $P^{a}$, $K^{i}$ and $J^{ij}$. But as subalgebras, both $Span\{P'^{a}, K'^{i},J'^{ij}\}$ and $Span\{P^{a}, K^{i},J^{ij}\}$ are isomorphic to $Gal(d-1,1)$.

At this point, one may ask what will happen if $\xi^{a}$ is taken to be spacelike $\langle\xi,\xi\rangle>0$ or timelike $\langle\xi,\xi\rangle<0$. In the former case we can perform a suitable Lorentz rotation so that $\xi^{a}$ is proportional to $(0,\cdots,0,1)$ and hence $\Pi\propto P^{d}$. The second constraint in (\ref{constraint}) then gives us the allowed solutions  $(\lambda^{a},\omega^{ab})$ as: $(\lambda^{a},\omega^{0i},\omega^{ij})=\hbox{free parameters}$, $\omega^{0 d}=\omega^{id}=0$. The centralizer $Z(\Pi)$ is thus spanned by $P^{0}$, $P^{i}$, $J^{0i}$ and $J^{ij}$, plus the central element $P^{d}$. Unlike the Galilei algebra, however, this center does not appear in the commutators of the remaining generators, thus completely decoupled, and we finally find that $Z(\Pi)$ is the Poincar\'e algebra $ISO(d-1,1)$ in $d$-dimensional spacetime (plus the decoupled center $P^{d}$). Evidently, the symmetry reduction $ISO(d,1)\rightarrow ISO(d-1,1)$ corresponds in field theory to the usual K-K compactification along the spatial direction $x^{d}$. 

If, on the other hand, $\xi^{a}$ is a timelike vector, a similar argument gives $Z(\Pi)\cong SO(d)\ltimes \mathbb{R}^{d}$, namely the Euclidean group in $d$ dimensions. In this connection, the continuous Wick rotation discussed in \cite{NW} can be interpreted as an $SO(2)$ rotation\footnote{Not the usual Lorentz boosts $SO(1,1)\subset SO(d,1)$ that preserve the causal structures.} in the $(\xi^{0},\xi^{d})$ plane, which turns a spacelike vector $\xi^{a}$ into a timelike one (or {\it vice versa}).
\subsubsection{Applying to the de Sitter Group}
Let us turn now to a similar analysis of the de Sitter group ${\cal G}_{d+1}=SO(d+1,1)$. When one works in the light-cone gauge $\xi^{0}=1$, $\xi^{i}=0$ and $\xi^{d}=-1$, the solutions of $A\in Z(\Pi)$ for $\Pi=P_{-}$ still take the form (\ref{solutions}), but now $\lambda^{a}$ are no longer free parameters; they must obey the first constraint in (\ref{constraint}): 
\begin{equation}
\lambda^{a}\xi^{b}-\lambda^{b}\xi^{b}=0\quad\Rightarrow\quad\left\{
\begin{array}{lll}
\lambda^{0}\xi^{i}-\lambda^{i}\xi^{0}=
\lambda^{i}\xi^{d}-\lambda^{d}\xi^{i}=0\quad
&\Rightarrow&\quad \lambda^{i}=0 \\
\lambda^{0}\xi^{d}-\lambda^{d}\xi^{0}=0 &\Rightarrow&  \lambda^{0}+\lambda^{d}=0
\end{array}\right.
\end{equation}
In other words, the only independent parameters here are $\lambda^{0}\equiv\frac{1}{\sqrt{2}}\lambda$, $\omega^{ij}$ and $\omega^{0i}\equiv -\frac{1}{\sqrt{2}}\omega^{i}$.
All the elements of $Z(\Pi)$ can then be expressed in terms of these free parameters:
\begin{equation}
A=\lambda P_{-}+\omega^{ij}J_{ij}+\omega^{i}K_{i}.
\end{equation}
As a consequence, the generalized light-cone subalgebra $G_{\rm DLCQ}$ of $SO(d+1,1)$ is generated by a center $P_{-}$, together with $\{J_{ij},K_{i}\}$. The commutation relations read:
\begin{eqnarray}
[P_{-},K_{i}]&=&[P_{-},J_{ij}]\;=\;0,\quad
[J_{ij},K_{k}]\;=\;i(\delta_{ik}K_{j}-\delta_{jk}K_{i}),\cr
[J_{ij},J_{kl}]&=&i(\delta_{ik}J_{jl}+\delta_{jl}J_{ik}-\delta_{il}J_{jk}-\delta_{jk}J_{il}),\quad [K_{i},K_{j}]\;=\;0.
\label{Euclidean}
\end{eqnarray}
Thus, we see explicitly that the number of non-central generators in $G_{\rm DLCQ}$ is $(d-1)+(d-1)(d-2)/2=d(d-1)/2$, less than $\dim G_{\rm IMF}=d(d+1)/2$.

Note that the algebra generated by $\{J_{ij},K_{i}\}$ is isomorphic to the Lie algebra of Euclidean group in $d-1$ dimensions, $Z(\Pi)\cong SO(d-1)\ltimes\mathbb{R}^{d-1}$, and the center $P_{-}$ is in fact decoupled from the Euclidean group, since the commutation relations among $J_{ij}$, $K_{i}$ do not depend on $P_{-}$. (In the flat space case, however, the center $P_{-}=-P^{+}$ couples nontrivially to the remaining generators of the Galilei algebra, as it appears in the the right hand side of (\ref{Gal+}).) In particular when $d=2+1$, there is no way to get a coupled central extension $[K_{i},K_{j}]=i\kappa\epsilon_{ij}$ in the above algebraic construction.

So it is possible to build the light-cone subalgebra $G_{\rm DLCQ}$ of $SO(d+1,1)$ by means of the centralizer of the longitudinal momentum $\Pi=P_{-}$. Just as in the flat space case, one may also consider some other gauges, for example, a time-like $\xi^{a}$ such as $(1,0,\cdots,0)$ or a space-like $\xi^{a}=(0,\cdots,0,1)$. The corresponding centralizers will give us some algebraic cousins of $G_{\rm DLCQ}$. In the former case we have $\Pi=P_{0}$, and solving the constraints yields the free parameters $\lambda^{0}$, $\omega^{di}$ and $\omega^{ij}$, so that $Z(\Pi)$ is generated by $\{J_{ij},J_{di}\}$ plus the decoupled center $P_{0}$, which leads to the rotation group $SO(d)\oplus\{P_{0}\}$ in $d$ dimensions. In the latter case where $\xi^{a}=(0,\cdots,0,1)$, we find $Z(\Pi)\cong Span\{J_{ij},J_{0i},P_{d}\}\cong SO(d-1,1)\oplus\{P_{d}\}$, and the central element $P_{d}$ is also decoupled. In either case the number of non-central generators is $d(d-1)/2$, the same as $\dim G_{\rm DLCQ}$.

The above results, though negative for seeking equivalence of IMF and DLCQ, may have some interesting consequences. Recall that in flat space, a continuous Wick rotation in $d$ dimensions can be interpreted from a $(d+1)$-dimensional perspective \cite{NW}, {\it i.e.}, as an $SO(2)$ rotation in the $(\xi^{0},\xi^{d})$ plane. When one rotates $\xi^{\mu}$ from a spacelike vector into some timelike one, the underlying spacetime symmetry goes from the Poincar\'e group $SO(d-1,1)\ltimes\mathbb{R}^{d-1,1}$ to the Euclidean group $SO(d)\ltimes\mathbb{R}^{d}$. The two phases are separated by the wall of null vectors $\xi^{a}$, on which the spacetime symmetries become the Galilei group $Gal(d-1,1)$. 

Now for the de Sitter case, we have a somewhat different (but quite similar) interpretation. Let us start with a field theory defined on $(d-1)$-dimensional de Sitter space $dS_{d-1}$, with the isometry group $SO(d-1,1)$. The above discussion indicates that this symmetry group can be embedded into the $(d+1)$-dimensional de Sitter group $SO(d+1,1)$ as the centralizer $Z(\Pi)$ of $\Pi=P_{d}$. This corresponds to the choice of a spacelike $\xi^{a}=(0,\cdots,0,1)$. We can rotate this vector into a timelike vector $\xi^{a}=(1,0,\cdots,0)$ and, accordingly, we find $\Pi=P_{0}$, so that the centralizer becomes $Z(\Pi)=SO(d)$, which is the isometry group of the manifold $S^{d-1}$, namely the Wick rotated version of $dS_{d-1}$. The wall separating $dS_{d-1}$ and $S^{d-1}$ consists of null vectors $\xi^{a}$, and for a null $\xi$, the centralizer $Z(\xi\cdot P)$ gives rise to the inhomogeneous Euclidean group $SO(d-1)\ltimes\mathbb{R}^{d-1}$ as the symmetry group. Thus, for de Sitter space, Wick rotations in $(d-1)$-dimensions may have a $(d+1)$-dimensional perspective. We believe that a similar argument should go through in anti de Sitter spaces, and this might provide a new way to look at the ``double Wick rotation'' \cite{bhm} $AdS_{p}\times S^{q}\rightarrow {\bf H}^{p}\times dS_{q}$, where ${\bf H}^{p}$ is the Euclidean version of $AdS_{p}$, namely, the $p$-dimensional hyperbolic space. 
\subsubsection{A Further Generalization}
As a further generalization of the concept of ``generalized light-cone gauge'', we extend (\ref{center}) by defining
\begin{equation}
\Pi=\sum_{a=0}^{d}\xi^{a}P_{a}+\frac{1}{2}\sum_{a,\,b=0}^{d}\zeta^{ab}J_{ab}
\label{center'}
\end{equation}
for some fixed $(\xi^{a},\zeta^{ab})$, with $\zeta^{ab}=-\zeta^{ba}$. This generalization is motivated by the fact that the momentum and angular momentum operators in the de Sitter algebra are more symmetric than those in the corresponding Poincar\'e algebra. In fact, let us introduce an index $A$ running from 0 to $d+1$, and define $J^{AB}$ with $J^{AB}=-J^{BA}$ by:
\begin{equation}
J^{AB}=\left\{\begin{array}{ll} J^{ab} & \hbox{if $A=a,B=b\in\{0,1,\cdots, d\}$;}\\
R P^{a}\quad & \hbox{if $A=a\in\{0,1,\cdots, d\}, B=d+1$.}
\end{array}\right.
\label{generator}
\end{equation}
In terms of these generators, the algebraic relations (\ref{deSitter+}) can be expressed more symmetrically
\begin{equation}
[J^{AB},J^{CD}]=i(\eta^{AC}J^{BD}+\eta^{BD}J^{AC}-\eta^{AD}J^{BC}-\eta^{BC}J^{AD}),
\label{deSitter'}
\end{equation}
where $\eta_{AB}$ denotes the Minkowski metric in $d+2$ dimensions,  $\eta_{AB}=diag(-++\cdots+)$. This shows explicitly that the de Sitter group in $(d+1)$-dimensions is isomorphic to $SO(d+1,1)$.

So let us rewrite (\ref{center'}) as
\begin{equation}
\Pi=\frac{1}{2}\sum_{A,\,B=0}^{d+1}\Xi^{AB}J_{AB}
\label{center"}
\end{equation}
with some fixed coefficients $\Xi^{AB}=-\Xi^{BA}$. In this new definition the ``light-cone like'' gauge condition $\Pi=\mu\cdot{\bf 1}$ is actually a gauge that fixes a mixed combination of the momentum and the angular momentum operators. Let us also rewrite the elements (\ref{expansion}) of $Z(\Pi)$ in the form
\begin{equation}
A=\frac{1}{2}\sum_{A,\,B}\Omega^{AB}J_{AB},\quad\Omega^{AB}=-\Omega^{BA}
\label{expansion'}
\end{equation}
and compute the commutator between (\ref{center"}) and (\ref{expansion'}). The result simply reads
\begin{equation}
[A,\Pi]=i\,\eta_{AC}\,\Omega^{AB}\,\Xi^{CD}J_{BD}.
\end{equation}
Thus, the condition of $A\in Z(\Pi)$ gives the constraints for the coefficients $\Omega^{AB}$ that appear in the expansion $A$:
\begin{equation}
\eta_{AC}(\Omega^{AB}\,\Xi^{CD}-\Omega^{AD}\,\Xi^{CB})=[\Xi,\,\Omega]^{BD}=0,
\label{constraint'}
\end{equation}
 where $[\Xi,\,\Omega]$ denotes the commutator between the $(d+2)\times (d+2)$ antisymmetric matrices $\Xi$ and $\Omega$.

As an example, we consider $\Pi=J_{\hat{A}\hat{B}}$ with fixed indices $\hat{A}\neq\hat{B}$. It amounts to the choice of $\Xi^{AB}=\delta^{[A}_{\hat{A}}\delta^{B]}_{\hat{B}}$ in (\ref{center"}). Substituting this into the constraints (\ref{constraint'}), one easily finds their solutions:
\begin{equation}
\Omega^{AB}=\left\{\begin{array}{lll}
\Omega^{A'B'}&=&\hbox{free parameters, if $A',B'\in\{0,1,\cdots,d+1\}-\{\hat{A},\hat{B}\}$},\cr
\Omega^{A\hat{A}}&=&\Omega^{A\hat{B}}\;=\;0,\cr
\Omega^{\hat{A}\hat{B}}&=&\hbox{free parameters}.
\end{array}\right.
\end{equation}
Thus, $Z(\Pi)$ is spanned by the generators $J_{A'B'}$ plus the decoupled center $J_{\hat{A}\hat{B}}$. This subalgebra is isomorphic either to $SO(d)$ or to $SO(d-1,1)$, depending on whether one of the indices $\hat{A},\hat{B}$ belongs to the time direction. The dimension of such a subalgebra is again $d(d-1)/2$, equal to the number of generators in $G_{\rm DLCQ}$ and less than $\dim G_{\rm IMF}$.
\section{The Newton-Hooke Model}\label{s3}
Now we turn to constructing our matrix model. This relies simply on an extension of single-particle dynamics. So in Section \ref{ss3.1} we discuss the ordinary one-particle systems that realize the Newton-Hooke symmetries, with and without the second central extension $\kappa$. This could be viewed as a study of the simplified, $1\times 1$ matrix model. In Section \ref{ss3.2}, we generalize this discussion by truely using matrix degrees of freedom, and explore some physical consequences.
\subsection{Single Particle Dynamics}{\label{ss3.1}
Let us consider first the case when the ``boundary'' spacetime dimension $d$ is not equal to $2+1$, or $d$ equals $2+1$ but the second central extension $\kappa$ vanishes. We start with the Lagrangian of a nonrelativistic massive particle in the presence of an inverted harmonic oscillator potential
\begin{equation}
{\cal L}_{0}=\frac{1}{2}\,\mu\,\dot{\bf x}^{2}+\frac{\mu}{2R^{2}}\,{\bf x}^{2},
\label{L_0}
\end{equation}
where ${\bf x}=(x^{1},\cdots,x^{d-1})=x^{i}$. The equations of motion simply read:
\begin{equation}
\ddot{\bf x}-\frac{1}{R^{2}}\,{\bf x}=0.
\label{EOM_0}
\end{equation}
Introducing the canonical momentum $p_{i}$ and the Hamiltonian $H=H({\bf x},{\bf p})$ via
\begin{equation}
p_{i}=\frac{\partial{\cal L}_{0}}{\partial \dot{x}^{i}}=\mu \dot{x}^{i}, \quad
H={\bf p}\cdot\dot{\bf x}-{\cal L}_{0}=\frac{1}{2\mu}{\bf p}^{2}-\frac{\mu}{2R^{2}}\,{\bf x}^{2},
\label{PE}
\end{equation}
we see that the equations of motion (\ref{EOM_0}) take the canonical form:
\begin{equation}
\dot{\bf x}=\{{\bf x},H\}_{\rm PB}=\frac{1}{\mu}\,{\bf p},\quad \dot{\bf p}=\{{\bf p},H\}_{\rm PB}=\frac{\mu}{R^{2}}\,{\bf x}, 
\label{Canonical_0}
\end{equation}
here $\{\cdot,\cdot\}_{\rm PB}$ denotes the Poisson bracket with $\{x^{i},p_{j}\}_{\rm PB}=\delta^{i}_{j}$ and $\{x^{i},x^{j}\}_{\rm PB}=\{p_{i},p_{j}\}_{\rm PB}=0$.

Now let us show how the Newton-Hooke symmetries manifest in the above simple system. Clearly, the Lagrangian (\ref{L_0}) is invariant under space rotations ${\bf x}\rightarrow{\cal R}\cdot{\bf x}$ and time translations $t\rightarrow t+b$, and we get two kinds of obvious Noether charges, {\it i.e.} the angular momentum $J^{ij}=x^{i}p^{j}-x^{j}p^{i}$ and the Hamiltonian $H$. On the other hand, under the Newton-Hooke boost $\delta_{\bf v}{\bf x}={\bf v}R\sinh(t/R)$ with a small velocity $\bf v$, the Lagrangian ${\cal L}_{0}$ is changed by an amount
\begin{equation}
\delta_{\bf v}{\cal L}_{0}=\mu\dot{\bf x}\cdot\delta_{\bf v}\dot{\bf x}+\frac{\mu}{R^{2}}\,{\bf x}\cdot\delta_{\bf v}{\bf x}=\mu{\bf v}\cdot\frac{d}{dt}\left[{\bf x}\,\cosh\left(\frac{t}{R}\right)\right],
\label{delta_v}
\end{equation}
which is a total derivative. Thus, using the standard Noether method, we find the following conserved charge
\begin{equation}
{\bf K}=\mu\,{\bf x}\cosh\left(\frac{t}{R}\right)-{\bf p}R\sinh\left(\frac{t}{R}\right)
\end{equation}
generating the Newton-Hooke boosts. That ${\bf K}$ is constant in time can be easily checked with the help of the Hamiltonian equations (\ref{Canonical_0}),
$\dot{\bf K}=(\frac{\mu}{R}\,{\bf x}-R\,\dot{\bf p})\sinh(\frac{t}{R})+(\mu\dot{\bf x}-{\bf p})\cosh(\frac{t}{R})=0$. Finally, from (\ref{NHtran}) we see that the infinitesimal space translation is given by $\delta_{\bf a}{\bf x}={\bf a}\cosh\frac{t}{R}$, where ${\bf a}$ is some small parameter. Such a transformation leads to the change in the Lagrangian
\begin{equation}
\delta_{\bf a}{\cal L}_{0}=\mu\dot{\bf x}\cdot\delta_{\bf a}\dot{\bf x}+\frac{\mu}{R^{2}}\,{\bf x}\cdot\delta_{\bf a}{\bf x}=\frac{\mu}{R}\,{\bf a}\cdot\frac{d}{dt}\left[{\bf x}\sinh\left(\frac{t}{R}\right)\right].
\label{delta_a}
\end{equation}
Since this is again a total derivative, the Noether theorem gives us another conserved charge
\begin{equation}
{\bf P}={\bf p}\cosh\left(\frac{t}{R}\right)-\frac{\mu{\bf x}}{R}\sinh\left(\frac{t}{R}\right).
\end{equation}
As a check, we find that
$\dot{\bf P}=(\dot{\bf p}-\frac{\mu}{R^{2}}\,{\bf x})\cosh(\frac{t}{R})+\frac{1}{R}({\bf p}-\mu\dot{\bf x})\sinh(\frac{t}{R})
$, which together with the canonical equations (\ref{Canonical_0}) shows that ${\bf P}$ is indeed a constant of motion.

We have seen that up to a total derivative term, the Lagrangian (\ref{L_0}) is invariant under the Newton-Hooke transformations (\ref{NHtran}). In particular, we have constructed the corresponding conserved charges, including the angular momentum $J^{ij}$, the Hamiltonian $H$, as well as the Newton-Hooke boost ${\bf K}$ and the conserved ``momentum'' ${\bf P}$. Since ${\bf P}(t)={\bf P}(0)={\bf p}(0)$, we find that ${\bf P}$ is nothing else than the canonical momentum defined in (\ref{PE}) at $t=0$. Similarly, we have ${\bf K}(t)={\bf K}(0)=\mu{\bf x}(0)$, so $\bf K$ is proportional to the particle position ${\bf x}$ at $t=0$. Now using the Poisson brackets between $x^{i}$ and $p^{j}$, it is an easy matter to derive the following relations:
\begin{eqnarray}
\{P^{i},P^{j}\}_{\rm PB}&=&0,\quad \{H,P^{i}\}_{\rm PB}=-\frac{1}{R^{2}}K^{i},\cr
\{J^{ij},P^{k}\}_{\rm PB}&=&\delta^{ik}P^{j}-\delta^{jk}P^{i},\quad \{J^{ij},H\}_{\rm PB}=0,\cr
\{J^{ij},J^{mn}\}_{\rm PB}&=& \delta^{im}J^{jn}+\delta^{jn}J^{im}-\delta^{in}J^{jm}-\delta^{jm}J^{in},\cr
\{P^{i},K^{j}\}_{\rm PB}&=&-\mu\,\delta^{ij},\quad \{H,K^{i}\}_{\rm PB}=-P^{i},\cr
\{J^{ij},K^{k}\}_{\rm PB}&=&\delta^{ik}K^{j}-\delta^{jk}K^{i},\quad \{K^{i},K^{j}\}_{\rm PB}\;=\;0.
\label{NH'}
\end{eqnarray}
Thus, after the canonical quantization $\{\cdot,\cdot\}_{\rm PB}\rightarrow [\cdot,\cdot]/i$, the above algebra is identical to the Newton-Hooke algebra (\ref{NH}), where the second central extension $\kappa$ does not appear. It gives us a clear understanding of how the symmetry group $NH(d-1,1)$ is realized in our dynamical system. Notice that both the ``momentum'' $P_{i}$ and the boost $K^{i}$ do not commute with the Hamiltonian: $[H,P_{i}]\neq 0$, $[H,K^{i}]\neq 0$. But this should not cause any trouble since, on the one hand, the operators $P_{i}$, $K^{i}$ depend {\it explicitly} on $t$ and, on the other hand, they are conserved quantities, in consistent with the correct evolution equations  $\frac{d}{dt}(\cdots)=i[H,(\cdots)]+\frac{\partial}{\partial t}(\cdots)$. 

Now, after quantization, the Hamiltonian $H$ defined in (\ref{PE}) becomes
\begin{equation}
H=-\frac{1}{2\mu}\sum_{i=1}^{d-1}\frac{\partial^{2}}{\partial x^{2}_{i}}-\frac{\mu}{2R^{2}}\sum_{i=1}^{d-1}x^{2}_{i}.
\label{qH}
\end{equation}
It describes a quantum mechanical particle of mass $\mu$ moving in $(d-1)$-dimensional space, with a potential $V({\bf x})=-\mu{\bf x}^{2}/(2R^{2})$. The spectrum of $H$ is clearly not bounded from below. This may be thought of as a nonrelativistic version of the fact that in de Sitter space, there does not exist conserved positive energy. One consequence of this fact is that the system does not allow a supersymmetric extension; otherwise we would have $H=\{Q,\bar{Q}\}\geq 0$. As another consequence, we see that a usual Newton-Hooke particle cannot be stable. To estimate its decay rate, recall that for the normal harmonic oscillator of frequency $\omega$, the wave function of a state at the energy level $E_{n}=(n+\frac{1}{2})\omega$ is $\psi_{n}(t)=\psi_{n}(0)e^{-i(n+\frac{1}{2})\omega t}$. We may continue $\omega$ to the imaginary frequency $\omega=\pm i/R$ so as to describe a Newton-Hooke particle. Its wave function then becomes $\psi_{n}(t)\sim\psi_{n}(0)e^{\pm nt/R}$, indicating that the lifetime is roughly $t\sim R$. Thus, although such a particle is unstable, it will live sufficiently long if $R$ is large.
\subsubsection{Incorporation of the Second Center}
We now work in $d=2+1$ dimensions and construct a single-particle system whose symmetry group is $NH(2,1)$, with both the first and the second central extensions $\mu$ and $\kappa$. Our model has the Lagrangian
\begin{equation}
{\cal L}=\frac{\mu}{2}\,\dot{x}_{i}^{2}+\frac{\mu}{2R^{2}}\,x_{i}^{2}-\frac{\kappa}{2}\,\epsilon_{ij}\left(\dot{x}_{i}\ddot{x}_{j}+\frac{1}{R^{2}}x_{i}\dot{x}_{j}\right)\equiv {\cal L}_{0}+{\cal L}_{\rm CS}
\label{L}
\end{equation}
where ${\cal L}_{0}$ is defined as in (\ref{L_0}), ${\cal L}_{\rm CS}$ is an additional term, which consists of two Chern-Simons like couplings --- one proportional to $\epsilon_{ij}\dot{x}_{i}\ddot{x}_{j}$ and the other proportional to $\epsilon_{ij}x_{i}\dot{x}_{j}$.

Our choice of the Lagrangian (\ref{L}) is based mainly on the following three requirements: (i) ${\cal L}_{\rm CS}$ is not a total derivative; thus its presence will have some nontrivial dynamical consequences, (ii) ${\cal L}_{\rm CS}$ is at most linearly dependent on $\ddot{x}_{i}$, so that the ghost problem arising from higher derivatives could be made harmless \cite{LSZ}, and (iii) adding ${\cal L}_{\rm CS}$ to (\ref{L_0}) does not violate the Newton-Hooke symmetries. These conditions fix ${\cal L}_{\rm CS}$ almost uniquely. 

To see (iii), one needs to check that under a small transformation (\ref{NHtran}), the change in $\cal L$ is at most a total derivative. We have already seen that ${\cal L}_{0}$ has such a property, so let us focus on the additional part ${\cal L}_{\rm CS}$. Thus, if one takes a space rotation $x_{i}\rightarrow x'_{i}={\cal R}_{ij}\cdot x_{j}$, then ${\cal L}_{\rm CS}$ will transform into ${\cal L}_{\rm CS}'=\det({\cal R}_{ij}){\cal L}_{\rm CS}$, which remains unchanged because ${\cal R}_{ij}\in SO(2)$ and therefore $\det({\cal R}_{ij})=1$. One may also consider a small boost $\delta_{v}x_{i}=v_{i}R\sinh(t/R)$, under which ${\cal L}_{\rm CS}$ is changed into ${\cal L}_{\rm CS}+\delta_{v}{\cal L}_{\rm CS}$, with
\begin{equation}
\delta_{v}{\cal L}_{\rm CS}=-\frac{\kappa}{2}\epsilon_{ij}\,v_{i}\frac{d}{dt}\left(\dot{x}_{j}\cosh\frac{t}{R}-\frac{1}{R}\,x_{j}\sinh\frac{t}{R}\right).
\label{delta_vCS}
\end{equation}
Hence $\delta_{v}{\cal L}_{\rm CS}$ is a total derivative. Similarly, when we perform a small transformation $\delta_{a}x_{i}=a_{i}\cosh(t/R)$, the change in ${\cal L}_{\rm CS}$ is given by
\begin{equation}
\delta_{a}{\cal L}_{\rm CS}=-\frac{\kappa a_{i}}{2R}\epsilon_{ij}\,\frac{d}{dt}\left(\dot{x}_{j}\sinh\frac{t}{R}-\frac{1}{R}\,x_{j}\cosh\frac{t}{R}\right),
\label{delta_aCS}
\end{equation}
which is again a total derivative. We thus find that (\ref{L}) indeed defines a model invariant under the Newton-Hooke transformations. We have written down the formulae (\ref{delta_vCS})-(\ref{delta_aCS}) explicitly since they will be useful when computing the corresponding Noether charges. 

The equations of motion may be derived from the generalized Euler-Lagrange equations. They take the form:
\begin{equation}
\kappa\,\epsilon_{ij}\stackrel{\dots}{x}_{j}-\mu\,\ddot{x}_{i}-\frac{\kappa}{R^{2}}\,\epsilon_{ij}\,\dot{x}_{j}+\frac{\mu}{R^{2}}\,x_{i}=0.
\label{EOM}
\end{equation}
Such equations can be put into a Hamiltonian form. To see this, let us apply the Ostrogradski formalism (see e.g. \cite{DIFF}) to the system containing higher order derivatives. We introduce the following two kinds of momenta
\begin{eqnarray}
p_{i}&\equiv&\frac{\partial{\cal L}}{\partial\dot{x}_{i}}-\frac{d}{dt}\frac{\partial{\cal L}}{\partial\ddot{x}_{i}}\;=\;\mu\,\dot{x}_{i}-\kappa\,\epsilon_{ij}\,\ddot{x}_{j}+\frac{\kappa}{2R^{2}}\,\epsilon_{ij}\,x_{j}\cr
\tilde{p}_{i}&\equiv&\frac{\partial{\cal L}}{\partial\ddot{x}_{i}}\;=\;\frac{\kappa}{2}\,\epsilon_{ij}\dot{x}_{j}
\label{pp}
\end{eqnarray}
which are canonically conjugate to $x_{i}$ and $\dot{x}_{i}$, respectively. Thus we get an eight-dimensional phase space $\Omega=\{x_{i},\dot{x}_{i},p_{i},\tilde{p}_{i}\}$, on which the Hamiltonian
\begin{equation}
H\equiv\dot{x}_{i}p_{i}+\ddot{x}_{i}\tilde{p}_{i}-{\cal L}=-\frac{2\mu}{\kappa^{2}}\,\tilde{p}_{i}^{2}-\frac{2}{\kappa}\,\epsilon_{ij}\,p_{i}\tilde{p}_{j}+\frac{1}{R^{2}}\,x_{i}\tilde{p}_{i}-\frac{\mu}{2R^{2}}\,x_{i}^{2}
\label{Hcan}
\end{equation}
is defined. Now, if this system were regular \cite{DIFF} or, equivalently, the eight coordinates of the phase space $\Omega$ were all independent, one could use the fundamental Poisson brackets
\begin{eqnarray}
\{x_{i},p_{j}\}_{\rm PB}&=&\{\dot{x}_{i},\tilde{p}_{j}\}_{\rm PB}\;=\;\delta_{ij}\cr
\{x_{i},\dot{x}_{j}\}_{\rm PB}&=&\{p_{i},\tilde{p}_{j}\}_{\rm PB}\;=\;\{x_{i},\tilde{p}_{j}\}_{\rm PB}\;=\;\{\dot{x}_{i},p_{j}\}_{\rm PB}\;=\;0\cr
\{x_{i},x_{j}\}_{\rm PB}&=&\{p_{i},p_{j}\}_{\rm PB}\;=\;\{\dot{x}_{i},\dot{x}_{j}\}_{\rm PB}\;=\;\{\tilde{p}_{i},\tilde{p}_{j}\}_{\rm PB}\;=\;0
\label{PB}
\end{eqnarray}
to rewrite the equations of motion (\ref{EOM}) in the canonical form of $\dot{f}=\{f,H\}_{\rm PB}+\partial_{t} f$, where $f$ is one of the coordinates $x_{i},\dot{x}_{i}, p_{i}$ and $\tilde{p}_{i}$, or a function in them (possibly depending explicitly on $t$).

Actually, however, our system is not regular and there exist two independent phase space constraints 
\begin{equation}
\varphi_{i}\equiv \dot{x}_{i}+\frac{2}{\kappa}\,\epsilon_{ij}\tilde{p}_{j}=0
\label{constraints}
\end{equation}
that arise from the second identity in (\ref{pp}). The existence of such constraints is a simple consequence of the fact that $\cal L$ has a linear dependence on $\ddot{x}_{i}$. The Poisson brackets between $\varphi_{i}$, $\varphi_{j}$ form a two by two matrix ${\cal C}_{ij}=\{\varphi_{i},\varphi_{j}\}_{\rm PB}=-\frac{4}{\kappa}\epsilon_{ij}$, whose inverse is given by
\begin{equation}
({\cal C}^{-1})_{ij}=\frac{\kappa}{4}\,\epsilon_{ij}.
\label{C}
\end{equation}
Following the standard procedure, we can then use the Dirac brackets
\begin{equation}
\{f,g\}_{\rm D}\equiv\{f,g\}_{\rm PB}-\{f,\varphi_{i}\}_{\rm PB}({\cal C}^{-1})_{ij}\{\varphi_{j},g\}_{\rm PB}
\label{Dirac}
\end{equation}
to define a symplectic structure on the 6-dimensional reduced phase space $\Omega_{\rm red}=\Omega/\{\varphi_{i}=0\}$. Since $\{x_{i},\varphi_{j}\}_{\rm PB}=\{p_{i},\varphi_{j}\}_{\rm PB}=0$, we find $\{x_{i},f\}_{\rm D}=\{x_{i},f\}_{\rm PB}$, $\{p_{i},f\}_{\rm D}=\{p_{i},f\}_{\rm PB}$ for any classical observable $f$ and, in particular, the coordinates $x_{i}$ themselves are commutative with respect to $\{\cdot,\cdot\}_{\rm D}$. We also have the following nontrivial Dirac brackets:
\begin{equation}
\{\dot{x}_{i},\dot{x}_{j}\}_{\rm D}=\frac{1}{\kappa}\,\epsilon_{ij},\quad
\{\tilde{p}_{i},\tilde{p}_{j}\}_{\rm D}=\frac{\kappa}{4}\,\epsilon_{ij},\quad
\{\dot{x}_{i},\tilde{p}_{j}\}_{\rm D}=\frac{1}{2}\,\delta_{ij}
\label{DB}
\end{equation}
together with $\{x_{i},p_{j}\}_{\rm D}=\{x_{i},p_{j}\}_{\rm PB}=\delta_{ij}$. All other Dirac brackets in the phase space vanish.

Now, with the help of the Dirac brackets, one easily checks that the equations of motion (\ref{EOM}) can be written canonically in the form
\begin{equation}
\frac{df}{dt}=\{f,H\}_{\rm D}+\frac{\partial f}{\partial t},
\label{can}
\end{equation} 
here $H$ is the Hamiltonian given in (\ref{Hcan}), and $f=f(x_{i},p_{i},\tilde{p}_{i};t)$ is an arbitrary function on the reduced phase space $\Omega_{\rm red}$, which may depend explicitly on $t$. Quantization of this system is quite straightforward: all one requires is to replace the Dirac brackets by the corresponding quantum mechanical commutators, $\{f,g\}_{\rm D}\rightarrow [f,g]/i$. Thus, as operators, we have:
\begin{equation}
[x_{i},p_{j}]=i\delta_{ij},\quad
[\tilde{p}_{i},\tilde{p}_{j}]=\frac{i\kappa}{4}\,\epsilon_{ij},\quad
[x_{i},x_{j}]=[p_{i},p_{j}]=[x_{i},\tilde{p}_{j}]=[p_{i},\tilde{p}_{j}]=0.
\label{pq}
\end{equation}
This shows that $p_{i}$ can be represented as the usual by the differential operators $p_{i}=-i\partial/\partial x_{i}$. Note that the operators  $\dot{x}_{i}$ can be solved from the constraints (\ref{constraints}) and they are related to $\tilde{p}_{i}$ via a simple linear combination. So in addition to (\ref{pq}), we also have $[\dot{x}_{i},\dot{x}_{j}]=i\epsilon_{ij}/\kappa$ and $[\dot{x}_{i},\tilde{p}_{j}]=i\delta_{ij}/2$. The quantized version of (\ref{can}) is then the Heisenberg equations of motion
\begin{equation}
\frac{df}{dt}=i\,[H,f]+\frac{\partial f}{\partial t}.
\label{Hei}
\end{equation}

It remains to see that the parameter $\kappa$ introduced in the Lagrangian (\ref{L}) is indeed the second center in the Newton-Hooke algebra (\ref{NH2+1'}). Let us apply now a generalized Noether method to the construction of conserved charges, and determine the algebraic relations they should obey. For a Lagrangian ${\cal L}={\cal L}(x_{i},\dot{x}_{i},\ddot{x}_{i})$ depending on the second order derivative of $x_{i}$, its variation with respect to $x_{i}\rightarrow x_{i}+\delta x_{i}$ is given, via the equations of motion, by
\begin{equation}
\delta{\cal L}=\frac{d}{dt}\,(p_{i}\,\delta x_{i}+\tilde{p}_{i}\,\delta\dot{x}_{i}).
\label{var1}
\end{equation}
This formula holds for arbitrary $\delta x_{i}$, and in particular for an infinitesimal symmetry transformation. On the other hand, if $\delta x_{i}$ is a symmetry transformation we have, without using the equations of motion,
\begin{equation}
\delta{\cal L}=\frac{d}{dt}\,(u_{i}\,\delta x_{i}).
\label{var2}
\end{equation}
The difference between (\ref{var1}) and (\ref{var2}) vanishes when the equations of motion are applied and therefore we get the following conserved charge
\begin{equation}
Q={\rm const.}\times (p_{i}\,\delta x_{i}+\tilde{p}_{i}\,\delta\dot{x}_{i}-u_{i}\,\delta x_{i}).
\label{Q}
\end{equation}
Now for the system defined by (\ref{L}), one observes that:

(i) The Lagrangian (\ref{L}) is invariant under the space rotation $\delta_{\omega} x_{i}=\omega_{ij}x_{j}=\omega\epsilon_{ij}x_{j}$, so one can set $u_{i}\delta_{\omega}x_{i}=0$ and hence the angular momentum
\begin{equation}
J=\epsilon_{ij}\,(x_{i}\,p_{j}+\dot{x}_{i}\,\tilde{p}_{j})=\epsilon_{ij}\,x_{i}\,p_{j}-\frac{2}{\kappa}\,(\tilde{p}_{i})^{2}
\label{J-k}
\end{equation}
is conserved. 

(ii) Under the Newton-Hooke boost $\delta_{v}x_{i}=v_{i}R\sinh(t/R)$, the change $\delta_{v}{\cal L}$ in (\ref{L}) is a total derivative. The quantity $u_{i}\delta_{v}x_{i}$ can be read off from a combination of (\ref{delta_v}) and (\ref{delta_vCS}). It follows that the boost operators
\begin{equation}
K_{i}=(\mu x_{i}-2\tilde{p}_{i})\cosh\frac{t}{R}+\left(\frac{\kappa}{2R}\,\epsilon_{ij}x_{j}-Rp_{i}\right)\sinh\frac{t}{R}
\label{K-k}
\end{equation}
are Noether charges. 

(iii) Similarly, under the symmetry transformation $\delta_{a}x_{i}=a_{i}\cosh(t/R)$, $\delta_{a}{\cal L}$ is also a total derivative. Thus, from (\ref{delta_a}), (\ref{delta_aCS}) and (\ref{Q}) we see that the ``momentum'' operators
\begin{equation}
P_{i}=\left(p_{i}-\frac{\kappa}{2R^{2}}\,\epsilon_{ij}x_{j}\right)\cosh\frac{t}{R}+\left(\frac{2}{R}\,\tilde{p}_{i}-\frac{\mu}{R}\,x_{i}\right)\sinh\frac{t}{R}
\label{P-k}
\end{equation}
are also conserved quantities. 

One may check that (\ref{J-k})--(\ref{P-k}) are indeed constant in time by applying the Heisenberg equations of motion (\ref{Hei}) together with the commutation relations (\ref{pq}). For example, one easily deduces from (\ref{Hcan}) and (\ref{pq}) that $[H,x_{i}]=2i\epsilon_{ij}\tilde{p}_{j}/\kappa$, $[H,p_{i}]=i(\tilde{p}_{i}-\mu x_{i})/R^{2}$, and $[H,\tilde{p}_{i}]=ip_{i}/2+i\epsilon_{ij}(\mu\tilde{p}_{j}/\kappa-\kappa x_{j}/(4R^{2}))$, so that the commutator between $H$ and (\ref{K-k}) gives
\begin{equation}
i\,[H,K_{i}]=\left(p_{i}-\frac{\kappa}{2R^{2}}\,\epsilon_{ij}x_{j}\right)\cosh\frac{t}{R}+\left(\frac{2}{R}\,\tilde{p}_{i}-\frac{\mu}{R}\,x_{i}\right)\sinh\frac{t}{R}=-\frac{\partial K_{i}}{\partial t}.
\end{equation}
This together with (\ref{Hei}) shows that $dK_{i}/dt=0$. It is also quite easy to check that $[H,K_{i}]=-iP_{i}$, $[K_{i},K_{j}]=i\kappa\epsilon_{ij}$ and so on. Thus, we finally find that the symmetry generators in our system can be realized by the Hamiltonian $H$, the angular momentum $J$, the Newton-Hooke boosts $K_{i}$, and the ``momentum'' operators $P_{i}$. They form the Newton-Hooke algebra (\ref{NH2+1'}) with the second center being precisely the coupling constant $\kappa$ in (\ref{L}).
\subsection{Matrix Quantum Mechanics}\label{ss3.2}
Now let us propose a matrix model that generalizes the 1-particle system explored in Section \ref{ss3.1}. We henceforth replace the $i$th coordinate $x^{i}$ in (\ref{L_0}) or (\ref{L}) by an $N\times N$ hermitian matrix $X^{i}$, $i=1,\cdots,d-1$. In the ``non-exotic'' case where $d\neq 2+1$, or $d=2+1$ but $\kappa=0$, this model has a Lagrangian looking like
\begin{equation}
{\cal L}={\rm Tr}\left\{\frac{\mu}{2}\,(\dot{X}^{i})^{2}+\frac{\mu}{2R^{2}}\,(X^{i})^{2}+\frac{1}{4g^{2}
l^{5}}\,[X^{i},X^{j}]^{2}+\cdots\right\}
\label{matrixL}
\end{equation}
here $g$ is a bulk coupling constant, $l$ a free length parameter introduced so as to get the correct dimension of the commutator term in (\ref{matrixL}) (which will be fixed in our later discussions, see (\ref{parameter})), and ``$\cdots$'' denotes some other terms, possibly including the fermionic contributions. This kind of matrix models, which contain a mass term $\frac{1}{2}M^{2}\,{\rm Tr}(X^{i})^{2}$, have been discussed previously in, e.g., \cite{TV}\cite{GY}\cite{kimura}, with completely different motivations.

The above matrix model is manifestly invariant under the Newton-Hooke transformations
\begin{equation}
X^{i}\rightarrow X^{i}+\left(v^{i}R\sinh\frac{t}{R}+a^{i}\cosh\frac{t}{R}\right){\bf 1}_{N\times N}.
\label{matrixNHtran}
\end{equation}
In fact, it is easy to see that such transformations will result in ${\cal L}_{0}\rightarrow {\cal L}_{0}+d(\cdots)/dt$. Notice that (\ref{matrixNHtran}) acts only on the center of mass of the system, $X^{i}_{\rm c.m.}$, and keeps the part $X^{i}_{\rm rel}$ of relative motion intact, where $X^{i}_{\rm c.m.}$ and $X^{i}_{\rm rel}$ are defined by
\begin{equation}
X^{i}_{\rm c.m.}=\frac{{\rm Tr}X^{i}}{N}\,{\bf 1}_{N\times N},\quad
X^{i}_{\rm rel}=X^{i}-X^{i}_{\rm c.m.}.
\end{equation}
Since $X^{i}_{\rm c.m.}$ is a multiple of the identity and $X^{i}_{\rm rel}$ is a traceless matrix, we have ${\rm Tr}(X^{i}_{\rm c.m.}X^{i}_{\rm rel})=[X^{i}_{\rm rel},(\cdots)]=0$, and therefore the Lagrangian (\ref{matrixL}) is decomposed into two decoupled pieces. The commutator terms appear only in the relative motion part of the Lagrangian.

Let us consider the fundamental classical objects described by (\ref{matrixL}) at rest. They could be roughly imagined as some bound states of KK excitations in the underlying bulk theory. In the flat space limit $R\rightarrow\infty$, the static equations of motion are simply $[X^{j},[X^{i},X^{j}]]=0$, which can be solved by mutually commuting matrices. One can diagonalize these matrices simultaneously to get a system of $N$ decoupled static D-particles \cite{BFSS}, or equivalently the KK excitations in D=11 M-theory. When $R$ has a finite value, however, the static equations of motion derived from (\ref{matrixL}) take the form
\begin{equation}
[X^{j},[X^{i},X^{j}]]+\frac{g^{2}l^{5}\mu}{R^{2}}X^{i}=0.
\label{static}
\end{equation}
Thus, for $d\geq 4$, we have the fuzzy sphere solution
\begin{equation}
X^{a}=\beta\cdot J^{a} \quad (a=1,2,3),\quad \beta^{2}\equiv\frac{g^{2}l^{5}\mu}{2R^{2}}
\label{fuzzy}
\end{equation}
where $J^{a}$ denotes a basis of $SU(2)$ generators in an $N$-dimensional representation, satisfying the commutation relations $[J^{a},J^{b}]=i\epsilon^{abc}J^{c}$. Note that this solution even applies to the case of $d=2+1$ if we neglect $X^{3}$ in (\ref{fuzzy}). The radius $L$ of this fuzzy sphere is determined by $\beta$ as well as the second Casimir $C_{2}(J)$ of the $N$-dimensional representation:
\begin{equation}
L^{2}=\beta^{2}\,C_{2}(J).
\label{radius}
\end{equation}
From (\ref{fuzzy})--(\ref{radius}) we see that $L$ shrinks when $R$ becomes large, if $g,l,\mu$ are held fixed. This could be viewed as a kind of the UV-IR relations since, on the boundary side, $L$ is the length scale within which a point can be no longer localized and, on the bulk side, $R$ is the horizon size for an inertial observer, and also the size of the spatial section of de Sitter space at $t=0$.

The static energy of this configuration therefore reads:
\begin{equation}
E=-\frac{g^{2}l^{5}\mu^{2}}{8R^{4}}\,NC_{2}(J).
\label{staticE}
\end{equation}
This clearly takes a smaller value than the energy of the trivial solution $X^{i}\equiv 0$. Moreover, given the matrix dimension $N$, a larger value of $C_{2}(J)$ produces a smaller energy, and when the $N$-dimensional representation becomes irreducible, (\ref{staticE}) reaches its lowest value
\begin{equation}
E_{\rm min}=-\frac{g^{2}l^{5}\mu^{2}}{32R^{4}}\,N(N^{2}-1).
\label{E_0}
\end{equation}
Accordingly, the fundamental classical objects in our model are spherical 2-branes of the size
\begin{equation}
L^{2}=\frac{g^{2}l^{5}\mu}{8R^{2}}\,(N^{2}-1),
\label{size}
\end{equation}
which are similar to the dielectric branes \cite{myers} formed by D-particles in the presence of higher rank RR background fields. Physically, however, there is a difference between the spherical 2-branes considered here and Myers' dielectric branes: In the dielectric case, the static equations of motion look like $[[X^{a},X^{b}],X^{b}]+if\epsilon_{abc}[X^{b},X^{c}]=0$, which is {\it not} invariant under the space inversion $X^{a}\rightarrow -X^{a}$ that takes the opposite orientation of a dielectric brane. This means that ``anti'' dielectric branes cannot exist \cite{GY}, reflecting the well-known fact that dielectric branes are neutral, carrying only dipole (or mutipole) moments and no RR charges \cite{myers}. But in our situation the static equations (\ref{static}) keep invariant under the space inversion. Thus, if $X^{a}$ is a spherical 2-brane solution (\ref{fuzzy}), then the ``anti'' brane $\tilde{X}^{a}=-X^{a}$ will give another solution. One could therefore prepare some system containing both a spherical brane ${\cal B}$ and its anti configuration $\bar{\cal B}$ here.

We now show that the matrix dimension, $N$, cannot be arbitrarily large, and give a rough estimate of its upper bound $N_{\rm max}$. Our argument follows McGreevy, Susskind and Toumbas \cite{MST} closely. At fixed time $t$, de Sitter space has a compact spatial section whose size in the metric (\ref{Gmetric}) is given by $R(t)=R\cosh(t/R)$ and, in particular, we have $R(t)=R$ at $t=0$. Now, if a static spherical 2-brane described in (\ref{fuzzy}) lives at a finite time $t$, it should also be alive at $t=0$. The size of this fuzzy sphere should be bounded by the size of its surrounding space: $L\leq R$. Thus, according to (\ref{size}), we should have
\begin{equation}
L^{2}\;\simeq\;\frac{g^{2}l^{5}\mu}{8R^{2}}\,N^{2}\;\leq\;R^{2}\quad\quad\Rightarrow\quad\quad N\;\leq\;N_{\rm max}\;\simeq\;\frac{1}{g}\,\sqrt{\frac{8}{l^{5}\mu}}\,R^{2}.
\label{Nbound}
\end{equation}
When $N$ reaches its upper bound $N_{\rm max}$, we can no longer add new degrees of freedom to the system so as to increase the matrix dimension.

The above estimation, though quite rough, may have two interesting implications. First, as stated by the holographic principle \cite{holo}, the number of degrees of freedom within a region of size $R$ for $(d+1)$-dimensional bulk gravity cannot exceed the upper bound $c\cdot R^{d-1}$, where the coefficient $c$ is essentially one over the Planck area. Thus, according to (\ref{Nbound}), our matrix model would not reproduce holography correctly for arbitrary $d$ except for $d=2+1$. When $d=2+1$, the supposed dual theory in the bulk is 4-dimensional, and only in this dimension (\ref{Nbound}) is consistent with the holographic principle, where $N_{\rm max}$ agrees with the Bekenstein-Hawking entropy in $dS_{4}$ up to some factor that could be fine-tuned. That our model looks plausible only in a special dimension is not that surprising. Recall that in flat background, the usual Matrix Quantum Mechanics can holographically describe bulk gravity only in 11 dimensions; gravity in other dimensions can still be described by M(atrix) theory via some T-duality, but the resulting theory does not constitute a usual quantum mechanical model, --- it should be something else like, for example, a Yang-Mills {\it field theory}. On similar grounds, we do not really expect that our matrix quantum mechanics can uniformly describe de Sitter gravity in general dimensions.

Second, if a single degree of freedom in our model could only have finitely many quantum states, namely $\dim{\cal H}<\infty$, where $\cal H$ is the one-particle Hilbert space, then the dimension of the total Hilbert space would be (neglecting subtleties arising from the particle statistics)
\begin{equation}
{\cal N}\simeq (\dim{\cal H})^{N_{\rm max}}<\infty.
\label{dim}
\end{equation}
In real life a quantum mechanical system described the Lagrangian (\ref{matrixL}) is unlikely to have a finite dimensional one-particle Hilbert space. For example, when $N=1$, such a system has the Hamiltonian (\ref{qH}) whose spectrum is unbounded from below. Let us forget about this hard problem for a moment and suppose that after some effort we can eventually construct a $\cal H$ (perhaps through a proper definition \cite{ew}) with $\dim{\cal H}<\infty$. With this assumption, the relation (\ref{dim}) will look quite appealing because: 

(i) The Bekenstein-Hawking entropy can be extracted from (\ref{dim}) by 
\begin{equation}
S\;\sim\;\log{\cal N},
\label{BHentropy}
\end{equation}
in accordance with the general expectation \cite{tb} \cite{ew}.

(ii) The dimension of the total Hilbert space grows exponentially like
\begin{equation}
{\cal N}\;\sim\;\exp\left(\frac{\log\dim{\cal H}}{g}\,\sqrt{\frac{8}{l^{5}\mu}}\,R^{2}\right)
\label{grow}
\end{equation}
as the coupling constant $g$ tends to zero. This behavior can be expressed in a more familiar form \cite{ew}, ${\cal N}\sim \exp(1/G^{2}\Lambda)$, where $G$ is Newton's gravitational constant in 4-dimension, and $\Lambda$ is the cosmological constant\footnote{Here we follow the convention adopted by Witten \cite{ew} to write an Einstein-Hilbert action. Therefore, our cosmological constant $\Lambda$ is related to the usual one $\Lambda_{0}\sim R^{-2}$ through $\Lambda=\Lambda_{0}/(8\pi G)$.}. Actually, a comparison between (\ref{Nbound}) and the Bekenstein-Hawking entropy formula suggests that $g\sim G/\sqrt{l^{5}\mu}$. We shall now adjust the free length $l$ such that
\begin{equation}
\sqrt{l^{5}\mu}\sim R^{2},\quad g\sim G^{2}\Lambda.
\label{parameter}
\end{equation}
This enables us to identify the dimensionless coupling constant $g$ in the bulk with $G^{2}\Lambda$.

The foregoing discussion shows that to make our model (at least qualitatively) reasonable for describing gravity in $dS_{4}$, we have to resolve the problem of $\dim{\cal H}=\infty$ or, equivalently, ${\cal N}=\infty$. We now try to take a few steps toward getting a possible dynamical solution of this problem.

As emphasized in \cite{ew}, the problem of ${\cal N}=\infty$ is closely related to the weak coupling limit of de Sitter gravity. In such a limit, the perturbation theory exhibits an infinite dimensional Hilbert space, which may correspond to certain unitary representation of the de Sitter group. When one departs from the weak coupling limit, however, the perturbation theory should break down at some point. This process is not expected to be continuous, namely it should not be a process in which the perturbation theory becomes worse and worse (as an asymptotic expansion) but never suddenly terminated. Otherwise, we will be unable to understand how the dimension of a Hilbert space can jump from infinity to a finite value. There must exist a discontinuity point. Unfortunately, up to now the bulk theory of gravity itself tells us almost nothing about what will happen at this discontinuity point.

Now it is quite natural to ask whether we can see a similar discontinuity in the matrix model (\ref{matrixL}). To address this question, we begin with the Hamiltonian of our model,
\begin{equation}
H={\rm Tr}\left\{\frac{1}{2\mu}\,\Pi_{i}^{2}-\frac{\mu}{2R^{2}}\,(X^{i})^{2}-\frac{\mu}{4g^{2}
R^{4}}\,[X^{i},X^{j}]^{2}+\cdots\right\}
\label{matrixH}
\end{equation}
where $\Pi_{i}$ is the momentum operator canonical to $X^{i}$. Clearly, the commutator term in (\ref{matrixH}) is non-negative for hermitian matrices, which will give a positive contribution to the energy as long as $X^{i}$ are not in the flat directions. To simplify our consideration, let us focus on a case when the Hamiltonian comes near to a special point of moduli space, $gR\approx 0$. At this point, all degrees of freedom in the non-flat directions are so costly in energy that they become completely frozen (this can be easily seen by considering the ratio of the commutator term to the mass term). The remaining degrees of freedom, which we still can see at small $gR$, are described by matrices along the flat directions. Such matrices are mutually commuting and thus the Hamiltonian (\ref{matrixH}) is decomposed into a sum of $d-1$ decoupled one-matrix model Hamiltonians, $H_{0}(X^{i})$, $i=1,\cdots,d-1$, with
\begin{equation}
H_{0}(X)={\rm Tr}\left\{-\frac{1}{2\mu}\left(\frac{\partial}{\partial X}\right)^{2}-\frac{\mu}{2R^{2}}X^{2}+\cdots\right\}.
\label{matrixH_0}
\end{equation}
In this way the model (\ref{matrixH}) reduces essentially to the matrix model of c = 1 strings \cite{c=1}. 

Note that in this small $gR$ sector, one cannot see the classical objects described in (\ref{fuzzy}), since the construction of these objects requires nonvanishing commutators between different matrix coordinates. This does not contrast with the earlier established fact that fuzzy sphere solutions are energetically favorable. Actually, according to (\ref{size}), we find that such fuzzy spheres will shrink to a point when $gR\rightarrow 0$, thus becoming the ordinary point degrees of freedom, with commutative spatial coordinates. It is also possible to see this directly from the static equations of motion (\ref{static}).

Now, one may have two ways to reach the special moduli, $gR\rightarrow 0$, at which our matrix model gets drastically simplified. One way is to keep $R$ very large but fixed, and let $g$ approach zero. This should correspond to the case where we consider the perturbation theory of gravity in large de Sitter space, or at a very low Hawking temperature $T_{\rm H}\sim 1/R\sim 0$. Another way is to hold $g$ away from vanishing and let $R$ approach zero. In that case, we are supposed to deal with the nonperturbation theory of gravity in small de Sitter space, or at an extremely high Hawking temperature. Both cases are described by the same c = 1 matrix model (\ref{matrixH_0}), but with a quite different parameter $R$ and at a quite different temperature $T_{\rm H}$. We henceforth expect that there may exist a kind of Kosterlitz-Thouless phase transition \cite{GK} that separates these two cases into two different phases. The low temperature phase takes place at $g\sim 0$ and $R\gg R_{c}$, corresponding to the perturbation theory of gravity in large de Sitter space. On the other hand, the high temperature phase may occur at a finite $g$ with $R\rightarrow 0$, which should describe the nonperturbation theory of gravity in very small de Sitter space.

Thus, there may exist a critical temperature $T_{c}\sim 1/R_{c}$, above which the system, which presumably corresponds to gravity in small de Sitter space and at strong coupling, goes into a phase of matrix chains describing c $<$ 1 strings. In this dual description the time variable is discretized with a wide lattice spacing \cite{GK}
\begin{equation}
\epsilon\;\sim\;\frac{1}{R}\;\sim\;T_{\rm H}. 
\label{UvIr}
\end{equation}
Note that (\ref{UvIr}) may be interpreted as another kind of the UV-IR relations. Obviously, $\epsilon$ is the least uncertainty of time in matrix quantum mechanics, and its existence will impose an upper bound $E_{\rm max}\sim 1/\epsilon$ on energies. On the other hand, the Hawking temperature $T_{\rm H}$ is the smallest bulk energy scale; anyway, a particle with energy $E<T_{\rm H}$ in de Sitter space will be immediately thermalized.

If we lower the temperature $T_{\rm H}$ to a value below the KT critical point $T_{c}$, and at the same time let $g$ approach zero so that $gR\rightarrow 0$, the system will enter into another phase, which is described by the matrix model Hamiltonian (\ref{matrixH_0}) with continuous time variable, at a nearly vanishing temperature. This phase should correspond to the perturbative regime of de Sitter gravity. Hence, we find a possible dual picture to see how a discontinuity can appear when the perturbation theory of de Sitter gravity is breaking down.

We now turn to the dimension $\cal N$ of the relevant Hilbert spaces in the above picture. Let us first consider the low temperature phase. Of course, in this phase we should have ${\cal N}=\infty$, as there should exist infinitely many eigenstates of the Hamiltonian (\ref{matrixH_0}). This is quite consistent with the fact \cite{ew} that perturbatively de Sitter Hilbert space has infinite dimension. Here, by collecting some well known results (see e.g. \cite{c=1} \cite{GK}), we give a brief description of the total Hilbert space ${\cal H}_{\rm tot}$ in this low temperature phase of our matrix model.

We starting by diagonalizing the hermitian matrix $X$ as $X=U^{\dag}\,{\rm diag}(x_{1},\cdots,x_{N})\,U$, with $U$ being a $SU(N)$ rotation. In terms of the eigenvalues $x_{i}$ as well as some angular variables, the Hamiltonian (\ref{matrixH_0}) reads
\begin{equation}
H_{0}=-\frac{1}{2\mu\Delta(x)}\sum_{i=1}^{N}\frac{\partial^{2}}{\partial x_{i}^{2}}\Delta(x)-\frac{\mu}{2R^{2}}\sum_{i=1}^{N}x_{i}^{2}+\sum_{i<j}\frac{\Pi_{ij}^{2}}{\mu (x_{i}-x_{j})^{2}}
\label{diagH}
\end{equation}
where $\Delta(x)=\sum_{i<j}(x_{i}-x_{j})$ denotes the Vandermonde determinant and $\Pi_{ij}$ are the hermitian generators of $SU(N)$. The Hilbert space ${\cal H}_{\rm tot}$ is spanned by eigenfunctions of (\ref{diagH}), including the singlet states that are invariant under $SU(N)$ rotations, as well as non-singlet states. For a singlet state, the wavefunction $\Psi$ depends only on $x_{i}$ and we can write $\Psi_{f}(x)=\Delta(x)\Psi(x)$, where $\Psi_{f}(x)$ obeys fermionic statistics. This allows us to reduce the $N$-body eigenequation $H_{0}\Psi=E\Psi$ to a one-body problem of fermions:
\begin{equation}
\left(-\frac{1}{2\mu}\frac{d^{2}}{dx^{2}}-\frac{\mu}{2R^{2}}x^{2}\right)\psi_{n}(x)=e_{n}\psi_{n}(x);\quad E=\sum_{i=1}^{N}e_{n_{i}},\quad\Psi_{f}=\bigwedge_{i=1}^{N}\psi_{n_{i}}
\label{fermion}
\end{equation}
where $n_{i}\neq n_{j}$ if $i\neq j$, according to Pauli's exclusion principle. Thus, we see that the singlet part of ${\cal H}_{\rm tot}$ has the structure $\bigwedge_{i=1}^{N}{\cal H}^{(i)}$, where ${\cal H}^{(i)}$ is the Hilbert space of a single fermionic degree of freedom, spanned by the wavefunctions $\psi_{n}(x_{i})$. This part of ${\cal H}_{\rm tot}$ is infinite dimensional since $\dim{\cal H}^{(i)}=\infty$. The non-singlet part of ${\cal H}_{\rm tot}$ is much more complicated \cite{GK} but it also contains infinitely many states.

Next, we consider the high temperature phase, in which $g=\hbox{finite}$, $R\rightarrow 0$. As mentioned, our system in this phase has a dual description in terms of matrix chains. The action of such a matrix chain may look like
\begin{equation}
S=\sum_{a=1}^{Q}\left\{\frac{1}{2\epsilon}{\rm Tr}\,(X_{a+1}-X_{a})^{2}+\epsilon\,{\rm Tr}\,W(X_{a})\right\}.
\label{chainS}
\end{equation}
At first sight, this system contains $QN^{2}$ degrees of freedom, but since $QN(N-1)$ of them are the angular part that are somehow decoupled and can be integrated out, we really have $QN$ eigenvalues as the dynamical variables. In particular, the quantum partition function is given by
\begin{equation}
Z(\beta)=\int\left(\prod_{a=1}^{Q}\prod_{i=1}^{N}dx^{i}_{a}\right)\Delta(x_{1})\Delta(x_{Q})\,\exp\left\{-\beta\sum_{a,i}\left[\frac{1}{2\epsilon}\,(x_{a+1}^{i}-x_{a}^{i})^{2}+\epsilon\,W(x_{a}^{i})\right]\right\}
\label{PF}
\end{equation}
Now if $Q$ is sufficiently large, the dimension $\cal N$ of the total Hilbert space ${\cal H}_{\rm tot}$ can be estimated by ${\cal N}=(\dim{\cal H})^{QN}$, where $\cal H$ is some Hilbert space for a single particle. Imagine that such a particle is described by ordinary quantum mechanics but one takes {\it both} a UV cutoff $\epsilon$ {\it and} an IR cutoff $Q\epsilon$. Now, as usual, the UV cutoff imposes an upper bound on the allowed energies, $|E|\leq E_{\rm max}=1/\epsilon$, while the IR cutoff makes the energy spectrum discrete, with the level separation at least by an amount $\Delta E=1/(Q\epsilon)$. Clearly, such a system allows only finitely many energy levels, and the total number of them is estimated by $\dim{\cal H}=E_{\rm max}/\Delta E=Q$. This finally gives the total Hilbert space dimension of the matrix chain:
\begin{equation}
{\cal N}\;\sim\;\exp(NQ\log Q).
\label{finiteN}
\end{equation}
Thus, (\ref{finiteN}) is a finite number provided $Q<\infty$.

So we indeed have a consistent picture to look at de Sitter gravity at $gR\approx 0$. In particular, it explains why in the perturbation theory de Sitter Hilbert space has infinite dimension but, nonperturbatively, it can become finitely dimensional. However, our arguments that lead to this picture are still not quite satisfactory: To get a finite $\cal N$ in the high temperature phase, we have to take a finite size $Q$ of the matrix chain. This is somewhat {\it ad hoc} since we know that matrix chains with $Q=\infty$ can also appear as certain high temperature phase in the KT transition, perhaps in a more natural way \cite{GK}. Thus, we need a better understanding of the physical difference between $Q<\infty$ and $Q=\infty$. Put it into another way: if our picture for $Q<\infty$ is correct, then what is the physics at $Q=\infty$ we have missed so far?

The missed point is, when taking the scaling limit $gR\rightarrow 0$ we have, in addition to the previous choices (i) $g=\hbox{finite},R\rightarrow 0$ and  (ii) $g\rightarrow 0,R=\hbox{(large but) fixed}$, a third choice: {\it i.e.}, both $R$ and $g$ approach zero. This should correspond to weakly coupled gravity in small de Sitter space, or at high temperature. Since this case is still described by the matrix model (\ref{matrixH_0}), we expect a KT phase transition here. Thus, if we start with weakly coupled gravity in large de Sitter space and raise the Hawking temperature while keeping $g$ still small, we will finally arrive at weakly coupled gravity in small de Sitter space, which is again described by a matrix chain. But since this still corresponds to the perturbation theory of de Sitter gravity, we must take $Q=\infty$ so that $\cal N$, computed by (\ref{finiteN}), also becomes divergent. To get strongly coupled gravity at this high temperature, one needs to shift the value of $Q$ away from $\infty$, so that the matrix chain becomes finitely long. Thus, our picture suggests that even in the perturbative regime of de Sitter gravity where $g$ is held small, there should exist a kind of discontinuity corresponding to the KT phase transition that turns a c = 1 matrix model to an infinitely long matrix chain, or {\it vice versa}. This discussion also suggests that the coupling constant $g$ should be related to $Q$ via some function $g=f(Q)$ vanishing at $Q=\infty$, e.g. $g\sim 1/Q$. Thus our picture, if consistent, incorporates automatically the issue \cite{ew} that in the nonperturbation theory of de Sitter gravity, the dimensionless coupling constant should be valued in a discrete set. Another possible reasoning leading to this discreteness was given in \cite{vol}.

The foregoing discussions apply only to the case of $gR\rightarrow 0$, where our original matrix model (\ref{matrixH}) reduces to a much simpler one, (\ref{matrixH_0}). In this approach, a complete description of de Sitter gravity should also include an understanding of the case when $gR$ is away from zero and becomes finite. This should correspond to strongly coupled gravity in large de Sitter space. Unfortunately, we have got few insights about this nontrivial and really interesting case so far. Hope that we could return to this problem in near future.
\section{Conclusions}\label{s4}
In conclusion, we have made a couple of observations about the possible spacetime symmetries that could appear in a matrix model description of de Sitter gravity. Our observations rest mainly on an analogy with M(atrix) theory in flat space. We found that for de Sitter space, in particular for $dS_{4}$, the Newton-Hooke symmetries are plausible. Based on such symmetries, we studied quantum mechanics of both a particle and a system consisting of matrix degrees of freedom. We discussed how the matrix model could solve some problems in de Sitter gravity.

Our matrix model described by the Lagrangian (\ref{matrixL}) is the simplest realization of the Newton-Hooke group in $d=2+1$, where only one central extension, $\mu$, is incorporated. As we have discussed in Section \ref{ss3.1}, it is also possible to incorporate the second center $\kappa$ into the quantum mechanical system. It would be interesting to see whether such an exotic extension is relevant to de Sitter gravity.

The matrix model considered in this paper is quite different from that proposed recently in \cite{ML}. On the one hand, we do not require from the outset that the dimension of the matrix model Hilbert space should be finite. Finite dimensions appear only dynamically. This gives us the possibility to address some interesting questions such as why in the perturbation theory of de Sitter gravity, one finds an infinite dimensional Hilbert space. On the other hand, our matrix model looks closer to the BFSS model in flat space. Thus, it would be easier to take the flat space limit $R\rightarrow\infty$ in our approach.
\vskip .3cm
Acknowledgements. I would like to thank Miao Li, Yong-shi Wu, and Zhong-xia Yang for stimulating discussions. I also wish to thank Ming-ming Lu for helpful conversations. 

\end{document}